\documentclass[USenglish,review]{elsarticle}
\biboptions{sort&compress}
\usepackage[utf8]{inputenc}
\usepackage[OT4]{fontenc}
\usepackage{amsmath,amssymb}
\usepackage{textcomp}
\usepackage{hyperref}

\usepackage{graphicx}
\usepackage{axodraw4j}
\usepackage{color}

\usepackage{setspace}

\newcommand{\mh}{\ensuremath{m_h}}

\newcommand{\nl}{\nonumber \\ }

\def\DRbar{\overline{\text{DR}}} 
\def\MSbar{\overline{\text{MS}}} 
\def\ttil{\tilde{t}} 
\def\XtMS{\widehat{X}_t}
\def\tb{t_\beta} 

\def\sinb{s_\beta}
\def\a{\alpha}

\def\l{\lambda}

\def\D{\Delta}

  \title{\huge A Review of Higgs Mass Calculations in Supersymmetric Models}

  \author[1]{Patrick Draper}
  \author[2,3]{Heidi Rzehak}
  \address[1]{University of California Santa Barbara,  Santa Barbara, CA USA}
  \address[2]{Albert-Ludwigs-Universität Freiburg,   Physikalisches Institut, 79104 Freiburg, Germany}
  \address[3]{$\it{CP}^3$ Origins \& Danish Institute for Advanced Study (DIAS), University of Southern Denmark, 5230 Odense
  M, Denmark}

\begin{document}

  \begin{abstract}
  { The discovery of the Higgs boson is both a milestone achievement for the Standard Model and an exciting probe of new physics beyond the SM. One of the most important properties of the Higgs is its mass, a number that has proven to be highly constraining for models of new physics, particularly those related to the electroweak hierarchy problem. Perhaps the most extensively studied examples are supersymmetric models, which, while capable of producing a 125 GeV Higgs boson with SM-like properties, do so in non-generic parts of their parameter spaces. We review the computation of the Higgs mass in the Minimal Supersymmetric Standard Model, in particular the large radiative corrections required to lift $m_h$ to 125 GeV and their calculation via Feynman-diagrammatic and effective field theory techniques. This review is intended as an entry point for readers new to the field, and as a summary of the current status, including the existing analytic calculations and publicly-available computer codes. }
  \end{abstract}



%
%


  \maketitle

\section{Introduction}

The discovery of a Higgs boson with properties broadly in agreement with Standard Model (SM) predictions is a major success of the first run of
the Large Hadron Collider (LHC).  From invariant mass peaks in the decay modes $h\rightarrow\gamma\gamma$ and $h\rightarrow ZZ\rightarrow 4\ell$, the mass of the new scalar is already known with remarkable precision:
\begin{align}
m_h=125.09\pm0.21~{\rm (stat)}\pm0.11~{\rm (syst)}~{\rm GeV}\;,
\label{eq:mh125}
\end{align}
as determined in a recent ATLAS+CMS combined analysis~\cite{Aad:2015zhl}.

In the SM, the Higgs boson mass is largely determined by its quartic self-coupling $\lambda$, evaluated near the electroweak scale. $\lambda$ is a free parameter, and therefore most of the interesting information it tells us about the SM concerns the renormalization group behavior of the theory at much higher energies~\cite{Buttazzo:2013uya}.

In contrast, $m_h$ is well-known to provide a sensitive probe of physics beyond the SM, particularly new physics associated with the electroweak hierarchy problem (EWHP). Composite Higgs models of different types make different predictions for ranges of the quartic coupling, and therefore for the Higgs mass.\footnote{For reviews, see~\cite{Bellazzini:2014yua} and references therein.} Likewise supersymmetric models, which will be our focus, make interesting predictions for $m_h$.

It has long been known that the Minimal Supersymmetric Standard Model (MSSM) can easily incorporate a SM-like lightest Higgs boson and that its tree-level mass is bounded from above by the mass of the $Z$ boson. This result is not incompatible with~(\ref{eq:mh125}) for two reasons. First, as was originally shown in~\cite{Haber:1990aw,Okada:1990vk,Ellis:1990nz}, the Higgs mass prediction in the MSSM is subject to radiative corrections that can significantly modify the tree-level result. As a byproduct, the Higgs mass becomes correlated with other parameters of the theory. Second, non-minimal supersymmetric models can introduce additional terms that contribute to the mass at tree level, through non-decoupling $F$- and $D$-terms~\cite{Batra:2003nj,Maloney:2004rc,Barbieri:2006bg,Maniatis:2009re,Ellwanger:2009dp}.\footnote{A hybrid possibility is that radiative corrections from new fields in non-minimal models can give significant contributions to $m_h$; see, for example,~\cite{Martin:2009bg}.}

If (\ref{eq:mh125}) is to be obtained in the MSSM, the radiative corrections must be large, and a very precise calculation is required to limit the theoretical uncertainty from uncomputed higher order corrections.  The payoff is that a prediction for \mh\ as a function of the superpartner masses, together with the experimental value, can serve as a guide to parameter regions in which SUSY might live. Furthermore, while it is possible to saturate~(\ref{eq:mh125}) with tree level effects in non-minimal supersymmetric models, it is often the case that some radiative corrections from MSSM fields are {\it also} required (see, e.g.~\cite{Hall:2011aa,Cheung:2012zq}). Therefore, it is of general importance to supersymmetric phenomenology to have a high-precision calculation of these effects.

In this article we provide an introductory review of the theoretical status of the Higgs mass in
the MSSM.  We start
with an introduction to the MSSM Higgs sector in
Section~\ref{sec:higgs-sector-mssm}. In Sections~\ref{sec:FD} and~\ref{sec:RGE}, we describe the computation of the leading quantum corrections to $m_h$ in two frameworks: the ``Feynman-Diagrammatic (FD)" or ``fixed-order" calculation, and the ``Renormalization Group (RG)" or ``Effective Field Theory (EFT)" calculation. In both sections we discuss subleading corrections and the current state-of-the-art. For numerical work, there
are a number of publicly available computer programs that calculate $m_h$ in different ways and with different levels of precision.  We review the public codes with the most sophisticated computations of $m_h$ in Section~\ref{sec:surv-publ-comp}, and comment on codes that compute radiative corrections in supersymmetric models beyond the MSSM.  
In Section~\ref{sec:concl} we summarize and conclude.

\section{The Tree-Level MSSM Higgs Sector}
\label{sec:higgs-sector-mssm}
We begin by briefly reviewing the Higgs sector of the MSSM at tree level and establishing notation and conventions. Some familiarity with supersymmetry is assumed. Excellent introductory reviews of both supersymmetry and the tree-level Higgs sector include~\cite{Martin:1997ns,Djouadi:2005gj}.

In the SM, one scalar Higgs doublet $H$ is sufficient to break electroweak symmetry and give masses to the quarks and leptons. To build the MSSM Higgs sector, we might begin by promoting $H$ to a chiral superfield, but it turns out this is not enough. The minimal supersymmetric model requires two Higgs doublet chiral superfields, $\hat H_u$ and $\hat H_d$, whose lowest components yield two scalar doublets. The extra chiral multiplet is necessary for two reasons.
\begin{itemize}
\item Holomorphy and gauge invariance of the superpotential. The quark and lepton masses arise from Yukawa couplings in the superpotential:
\begin{align}
W\supset -h_u \hat H_u \hat Q\hat{\bar u}+h_d\hat H_d\hat Q\hat{\bar d}+h_e\hat H_d\hat L\hat{\bar e}\;,
\label{eq:yuka}
\end{align}
where $\hat Q$,  $\hat{\bar u}$, $\hat{\bar d}$, $\hat L$, $\hat{\bar e}$ denote superfields containing the $SU(2)_L$-doublet quarks, the up and down type singlet quarks, the doublet leptons, and the singlet charged leptons, respectively.
Gauge invariance requires that the Higgs fields coupling to $\hat Q\hat {\bar u}$ and $\hat Q\hat{ \bar d}$ have opposite hypercharge, while holomorphy requires that the fields are both chiral. In the SM, there is no holomorphy requirement, so the masses of the up-type quarks may be obtained from the conjugate of the field providing masses to the down-type fermions.
\item
Anomaly cancellation. Gauge anomalies cancel in the Standard Model, but promoting the Higgs doublets to superfields in the MSSM introduces new chiral fermions, the Higgsinos. Cancellation of the $SU(2)_L^2U(1)_Y$ and $U(1)_Y^3$ anomalies is maintained because the hypercharges of $\hat H_u$ and $\hat H_d$ are opposite.
\end{itemize}

In addition to the Yukawa couplings~(\ref{eq:yuka}), there is one other gauge-invariant holomorphic term we can include in the superpotential:
\begin{align}
W\supset \mu \hat H_u \hat H_d\;,
\end{align}
which gives mass to the Higgsinos, provides quadratic terms in the Higgs potential, and contributes to trilinear scalar interactions. 

The Higgs kinetic terms, gauge interactions, and scalar quartic interactions arise from the K\"ahler $D$-terms,
\begin{align}
\mathcal{L}_{\text{vector}} &= \bigl[ \hat{H}_u^\dagger e^{2 g' \hat V' + 2 g \hat V} \hat H_u+\hat{H}_d^\dagger e^{2 g' \hat V' + 2 g\hat V} \hat H_d\bigr]\bigl |_{\theta
  \theta \bar{\theta} \bar{\theta}}\;,
  \label{eq:Kahler}
\end{align}
where $\hat V \equiv T^a \hat{V}^a,~ \hat{V}' \equiv \frac{Y}{2} \hat{v}'$, and $\hat{V}^a$ and $\hat{v}'$ are the $SU(2)_L$ and $U(1)_Y$ gauge superfields, respectively, with gauge couplings $g$ and $g'$. 

In a model with exact supersymmetry, each fermionic degree of freedom is present with a mass-degenerate bosonic partner, and vice versa. Since mass-degenerate partners for the Standard Model fields have not been found, realistic supersymmetric models must incorporate some controlled amount of supersymmetry breaking. In the MSSM, the breaking is parametrized by ``soft" (dimensionful) terms in the action:
 \begin{align}
\begin{split}
{\mathcal{L}}_{\text{soft}} = &- m_{\tilde{Q}}^2 \tilde{Q}^\dagger\tilde{Q} - m_{\tilde{u}}^2 \tilde{\bar u}^\dagger\tilde{\bar u}- m_{\tilde{d}}^2 \tilde{\bar d}^\dagger\tilde{\bar d}- m_{\tilde{L}}^2 \tilde{L}^\dagger\tilde{L}- m_{\tilde{e}}^2 \tilde{\bar e}^\dagger\tilde{\bar e}
\\& - m_{H_d}^2 {{H}}_d^\dagger{{H}}_d
- m_{H_u}^2 {{H}}_u^\dagger{{H}}_u + (B_\mu H_uH_d +h.c.)\\&
+(h_u A_u H_u \tilde{Q}\tilde{\bar u}+h_d A_dH_d\tilde{Q}\tilde{\bar d}+h_e A_eH_d\tilde{L}\tilde{\bar e}+h.c.)\\&
+ \frac{1}{2}(M_1 {\lambda}_B {\lambda}_B + M_2 {\lambda}_W^a {\lambda}_W^a + M_3 {\lambda}_g^a {\lambda}_g^a + h.c.). 
\end{split}
\label{eq:Lsoft}
\end{align}
The first line provides soft breaking masses $m_{\tilde{Q}}^2$, $m_{\tilde{u}}^2$, etc. to the sfermions; the second line gives soft masses to the Higgs bosons; the third line contains soft trilinear Higgs-sfermion-sfermion interactions with dimension-1 ``$A$-term" couplings; and the fourth line provides soft masses for the bino, wino, and gluino. The soft sfermion masses and trilinear couplings are in general matrices in flavor space, but the absence of 
flavor-changing neutral currents suggests that either the sfermions are very heavy (perhaps a thousand times the TeV scale~\cite{Wells:2004di}), or the flavor structure is not random. A common hypothesis is that the soft mass matrices are approximately proportional to the unit matrix and that the trilinear couplings are proportional to the Yukawa couplings (as already indicated in Eq.~(\ref{eq:Lsoft})). A detailed specification of the flavor structure will not be essential to understand the dominant radiative corrections to the Higgs sector discussed in this review. For our purposes, since a SM-like Higgs boson couples most strongly to the top sector, the most important parameters in the soft Lagrangian will be masses of the stop squarks and their trilinear couplings $h_t A_t H_u \tilde{t}\tilde{\bar t}$.

The soft Lagrangian, the K\"ahler $D$-terms, and the superpotential all contribute to the scalar Higgs potential:
\begin{align}\nonumber
V_{\text{Higgs}}  &=  \frac{g^2 + {g'}^2}{8} (H_d^\dagger H_d
  - H_u^\dagger H_u)^2 + \frac{{g}^2}{2} |H_d^\dagger H_u|^2
+  |{\mu}|^2 (H_d^\dagger H_d + H_u^\dagger H_u) \quad \quad
\\[0.3cm]& \quad\
 + (m_{H_d}^2 H_d^\dagger H_d + m_{H_u}^2
H_u^\dagger H_u) 
-(B_\mu H_u H_d + h.c.)
\label{eq:Higgspot}
\end{align}
The quartic couplings arise from the $D$-terms and are thus constrained to be functions of the weak gauge couplings. The second line of \eqref{eq:Higgspot} contains the only complex parameter in the Higgs potential,
$B_\mu$. The phase of $B_\mu$ can be rotated away with a
Peccei--Quinn transformation, and we will see below that the minimum does not spontaneously break $CP$, so the Higgs sector is $CP$-conserving at tree-level.

Let us now analyze the vacuum structure of Eq.~\eqref{eq:Higgspot}. We express the scalar doublets in terms
of charged complex fields $\phi^+_u$, $\phi^-_d$, neutral real fields
$\phi_u$, $\phi_d$, $\zeta_u$, $\zeta_d$, and vacuum expectation values $v_d$ and $e^{i\varphi_u}v_u$ for the neutral components:
\begin{align}\label{eq:Higgsdoublets}
{H_d} = \begin{pmatrix} v_d + \frac{1}{\sqrt{2}}({\phi}_d - i
{\zeta}_d) \\ -{\phi}_d^{-} \end{pmatrix}~, \qquad\
{H_u} = e^{i \varphi_u} \begin{pmatrix} {\phi}_u^+ \\  v_u +
  \frac{1}{\sqrt{2}}({\phi}_u + i {\zeta}_u) \end{pmatrix}
\end{align}
One of the vacuum expectation values, which we choose to be $v_d$, can be made real with a hypercharge rotation, while we parametrize a possible phase difference between the vacuum expectation values with the angle
$\varphi_u$. The $Z$ boson mass arises from the couplings in Eq.~\eqref{eq:Kahler},
\begin{align}
M_Z^2=(g^2+g'^2)(|v_u|^2+|v_d|^2)/2\;,
\end{align}
and determines the combination $\sqrt{|v_u|^2+|v_d|^2}\approx 174$ GeV.
 
The vacuum conditions are determined by the vanishing of tadpoles $t_\Phi$:
\begin{align}
\frac{\partial V_{\text{Higgs}}}{\partial \Phi}|_{\Phi = 0} \equiv -t_\Phi= 0~,\quad \Phi
={\phi_u, \phi_d, \zeta_u, \zeta_d}.
\end{align}
At tree-level, these conditions read 
\begin{align}
\label{eq:tphiu}
0 &= -t_{\phi_u} = -\sqrt{2}
   \, B_\mu \,v_d \cos(\varphi_u) + \sqrt{2}m_2^2 v_u - \frac{ G^2 v_u (v_d^2
    -v_u^2)} {2 \sqrt{2}}~, \\
\label{eq:tphid}
0 &= - t_{\phi_d} 
=  -  \sqrt{2}\, B_\mu\, v_u\cos(\varphi_u) +  \sqrt{2}m_1^2 v_d+ \frac{
    G^2 v_d (v_d^2 -v_u^2)}{2 \sqrt{2}}~, \\
\label{eq:tzetau}
  0&= - t_{\zeta_u} = (v_d/v_u) t_{\zeta_d} = 
  \sqrt{2} \, B_\mu \, v_d \sin(\varphi_u)~. 
\end{align}
where we have introduced the notation $m_{1,2}^2\equiv m_{H_{d,u}}^2+|\mu|^2$ and $G^2\equiv g^2+{g'}^2$. The minimum conditions will change in the presence of radiative corrections; the introduction of the tadpole
parameters will be useful in taking these into account. It is most convenient to solve Eqs.~\eqref{eq:tphiu}--\eqref{eq:tphid} for $m_1^2$ and $m_2^2$ and set $\varphi_u=0$.

The real and complex scalar fields introduced in Eq.~\eqref{eq:Higgsdoublets}
describe the different interaction behavior of the Higgs doublet components,
but they are not mass eigenstates. The fields with the same quantum
numbers mix through the bilinear terms in the potential:
\begin{align}\nonumber
\hspace*{-0.1cm} V_{\text{Higgs}}|_{\text{bil.}} &= 
\frac{1}{2} \begin{pmatrix} \phi_d,  \phi_u \end{pmatrix}
 \mathcal{M}_{\phi}
\begin{pmatrix} \phi_d \\ \phi_u \end{pmatrix}
+ \frac{1}{2} \begin{pmatrix} \phi_d,  \phi_u \end{pmatrix}
 \mathcal{M}_{\phi\zeta}
\begin{pmatrix} \zeta_d \\ \zeta_u \end{pmatrix} 
+ \frac{1}{2} \begin{pmatrix} \zeta_d,  \zeta_u \end{pmatrix}
 \mathcal{M}_{\phi\zeta}^T
\begin{pmatrix} \phi_d \\ \phi_u \end{pmatrix} 
\\[0.1cm]& \quad
+ \frac{1}{2} \begin{pmatrix} \zeta_d,  \zeta_u \end{pmatrix}
 \mathcal{M}_{\zeta}
\begin{pmatrix} \zeta_d \\ \zeta_u \end{pmatrix} +
\begin{pmatrix} {\phi^+_d},  {\phi}^+_u \end{pmatrix}
 \mathcal{M}_{{\phi}^{\pm}}
\begin{pmatrix} {\phi}^{-}_d \\ {{\phi}^{-}_u} \end{pmatrix}~.
\end{align}
Before applying the minimization conditions Eqs.~\eqref{eq:tphiu}--\eqref{eq:tzetau}, the mass matrices for the neutral components read:
\begin{align}
\mathcal{M}_{\phi} = \begin{pmatrix} m_1^2 +
  \frac{1}{4}G^2 (3 v_d^2 - v_u^2) & -\bigl(B_\mu
  \cos( \varphi_u) 
 + \frac{1}{2} G^2 v_d v_u \bigr)\\[0.1cm] -\bigl(B_\mu
 \cos( \varphi_u) +
 \frac{1}{2} G^2  v_d 
  v_u\bigr) & m_2^2 -\frac{1}{4} G^2 (v_d^2 - 3
  v_u^2)
\end{pmatrix}~,\label{eq:Mphi}\\
\label{eq:Mzeta}
\mathcal{M}_{\zeta} = \begin{pmatrix}m_1^2 +
  \frac{1}{4}G^2
  (v_d^2 - v_u^2) & -B_\mu \cos(\varphi_u) \\[0.1cm] - B_\mu
\cos( \varphi_u) & m_2^2 -\frac{1}{4} G^2 (v_d^2 - v_u^2)
\end{pmatrix}~,
\end{align}
and
\begin{align} \label{eq:Mphizeta}
\mathcal{M}_{\phi\zeta} =  \begin{pmatrix}0 &B_\mu \sin( 
  \varphi_u) \\[0.1cm] - B_\mu \sin( \varphi_u) & 0\end{pmatrix}~.
\end{align}
Due to the condition \eqref{eq:tzetau}, the mixing $\mathcal{M}_{\phi\zeta}$
between $\zeta$ and $\phi$ fields vanishes at tree-level. Therefore, there
is no CP-violation in the MSSM Higgs sector at tree-level, and we can consistently refer to the $\zeta$ and $\phi$ fields and CP-odd and CP-even, respectively. At higher orders in perturbation theory, CP-violation can enter the Higgs sector.

Similarly, before minimization the mass matrix of the charged Higgs bosons
$\mathcal{M}_{\phi^{\pm}}$ reads:
\begin{align}\label{eq:Mphicharged}
\mathcal{M}_{\phi^{\pm}} = \begin{pmatrix} m_1^2 +
  \frac{1}{4}\bigl(G^2 v_d^2 + \tilde{G}^2 v_u^2\bigr) & - B_\mu \, e^{-i\varphi_u} -
  \frac{1}{2} g^2 v_d 
  v_u \\[0.1cm] - B_\mu \, e^{i\varphi_u} - \frac{1}{2} g^2 v_d v_u &
  m_2^2 +  
  \frac{1}{4}\bigl(\tilde{G}^2 v_d^2 +G^2 v_u^2\bigr)~ 
\end{pmatrix}
\end{align}
where  $\tilde{G}^2 \equiv g^2 - {g'}^2$.

Imposing the minimization conditions and diagonalizing the mass matrices yields the masses and the corresponding
mass 
eigenstates. The transformation from the interaction eigenstates $\phi_u$,
$\phi_d$, $\zeta_u$, $\zeta_d$, $\phi_u^\pm$, $\phi_d^\pm$ to the mass
eigenstates $h$, $H$, $G$, $A$, $G^\pm$, $H^\pm$ can be described by the unitary
mixing matrices $\mathcal U_n$ and $\mathcal U_c$, with
\begin{align}\label{eq:fieldtreetrafo}
 (h, H, A,  G)^T = \mathcal U_n (\phi_d, \phi_u, \zeta_d, \zeta_u)^T, \quad 
 (H^+, G^+)^T = \mathcal U_c ({\phi_d^+}, \phi_u^+)\;.
\end{align}
At tree level, the matrix $\mathcal U_n$ is block-diagonal:
\begin{align}
\mathcal U_n = \begin{pmatrix} U_\alpha & 0\\
                               0 & U_{\beta_m}
                   \end{pmatrix}
\end{align}
and the matrix $ \mathcal U_c =
U_{\beta_m}$. The matrices $U_\gamma$ take the form 
\begin{align}U_\gamma = \begin{pmatrix}
- \sin \gamma & \cos \gamma \\ \cos \gamma & \sin \gamma \end{pmatrix}, \quad
\gamma = \alpha, \beta_m\;.
\end{align}  
It is customary to define an angle $\beta$ by the ratio of the vacuum expectation values, $\tan\beta\equiv v_u/v_d$. 
At tree-level,  the mixing angle $\beta_m$ is identified with the angle
$\beta$, 
$\tan \beta_m = \tan \beta $. The mixing angle $\alpha$ satisfies
\begin{align}\tan 2 \alpha = 
\tan 2 \beta \frac{ B_\mu (\tan \beta +\cot \beta) +M_Z^2}{B_\mu
  (\tan \beta +\cot \beta) - M_Z^2}\;,
\end{align}
and for $M_A>M_Z$, $\alpha$ can be taken in the range $-\pi/2<\alpha<0$.
It should be
noted, however, that these relations can be changed at higher orders, in
particular, the mixing angle $\beta_m$ and $\beta$ might differ (and the mixing angle diagonalizing the charged Higgs mass matrix can also differ from $\beta_m$). 

 The tree-level masses of the CP-even Higgs bosons
$M^{(0)}_h$, $M^{(0)}_H$, are given by
\begin{align}\nonumber
\left(M^{(0)}_{H, h}\right)^2 &= \frac{1}{2}\biggl(B_\mu (\tan \beta +\cot
\beta) + M_Z^2 
\\ &\quad\ \pm \sqrt{\bigl(B_\mu (\tan \beta - \cot
  \beta) + M_Z^2 \cos(2 \beta)\bigr)^2 + \bigl(2 B_\mu  + M_Z^2 \sin(2
  \beta)\bigr)^2}\biggr)~,\label{eq:treemassHh}
\end{align} 
where $M^{(0)}_h$ corresponds to the minus sign.
The CP-odd Higgs
boson masses are
\begin{align}
\left(M^{(0)}_{G, A}\right)^2 
&= \{0, B_\mu (\tan \beta +\cot \beta)\}~, \label{eq:treemassGA}
\end{align}
and the charged Higgs masses are
\begin{align}
\left(M^{(0)}_{G^{\pm}, H^{\pm}}\right)^2
&= \{0, B_\mu (\tan \beta +\cot \beta) +M_W^2\}~.\label{eq:treemassGH}
\end{align}
In each case the vanishing masses correspond to the neutral and charged Goldstone
bosons (where gauge-fixing 
terms have not yet been taken into account). The nonzero masses 
$M_A$ and $M_{H^\pm}$ correspond to the physical CP-odd and
charged Higgs bosons of the MSSM. Since these masses are determined at tree level by the input parameters $B_\mu$ and $\tan\beta$, it is customary to exchange $B_\mu$ for either  $M_A$ or
$M_{H^\pm}$. In the CP-conserving MSSM, $M_A$ is more
commonly taken as input parameter, while in the CP-violating case
the charged Higgs boson mass is more useful (since the distinction between CP-even and CP-odd is not sharp once radiative corrections are included).

Replacing $B_\mu$ by the appropriate $M_A$ dependence in Eq.~\eqref{eq:treemassHh}, the lightest tree-level CP-even Higgs mass becomes
\begin{align}
\left(M_h^{(0)}\right)^2=\frac{1}{2}\left(M_A^2+M_Z^2-\sqrt{(M_A^2-M_Z^2)^2+4M_A^2M_Z^2\sin^2(2\beta)}\right)\;.
\label{eq:mh02}
\end{align}
Perhaps the feature of the MSSM Higgs sector that has generated the most attention and interest is that Eq.~\eqref{eq:mh02} is bounded from above,
\begin{align}
M_h^{(0)}< M_Z\;.
\label{eq:mhbound}
\end{align} 
The bound is saturated for large $M_A$ and large $\tan\beta$. In the ``decoupling limit" $M_A\gg M_Z$, the tree-level mass is $M_h^{(0)}=M_Z|\cos(2\beta)|$, and $h$ has SM-like couplings to the electroweak gauge bosons and fermions.
Although the SM-like couplings for $h$ obtained in the decoupling limit are consistent with the properties of the Higgs boson observed at the LHC, the observed mass of 125 GeV is much greater than $M_Z$. Either large tree-level corrections or large radiative corrections are needed to increase the upper bound~\eqref{eq:mhbound}. The former requires new field content beyond the MSSM. The latter can occur in the MSSM alone, and will be the subject of the following sections.

\section{Radiative Corrections: Feynman Diagrammatic Approach}
\label{sec:FD}
The loop-corrected Higgs-mass spectrum is given by the real part of the zeroes
of the determinant of the renormalized two-point vertex function $\hat{\Gamma}$:
\begin{align}\label{eq:2ptvtxfct}
- i \hat{\Gamma}(p^2) = p^2 - \mathcal M(p^2)
\end{align} 
where $\mathcal M(p^2)$ denotes the loop-corrected Higgs mass matrix
\footnote{We could also work in
 the interaction eigenstate basis, which, depending on the organization of the calculation, 
might be more convenient for calculating the self energies. On the other hand, to obtain the one-loop corrections to the lightest Higgs boson mass, only the self energy $\hat{\Sigma}_{hh}$ is needed. In this case, it is easier to
directly calculate in the basis of the tree-level mass eigenstates. If Higgs
bosons appear as internal particles in loop diagrams, then it is also simpler to
use tree-level mass eigenstates.}  
 with
entries
\begin{align}\label{eq:loopcorrHmassmatrix}
\mathcal M_{ij}(p^2) = (M_{H_i}^{(0)})^2 \delta_{ij} -
\hat{\Sigma}_{H_iH_j}(p^2) 
\end{align}
with $H_i = h, H, A, G, H^\pm, G^\pm$. The scalar-gauge boson mixing need not to
be taken into account, as the zeroes of
Eq.~\eqref{eq:2ptvtxfct} are the same as those of the extended propagator matrix~\cite{Hollik:2002mv}.  
The matrix $\mathcal M_{ij}(p^2)$ is block-diagonal with one
block for the neutral and one for the charged Higgs bosons. In the case of
CP-invariance, the loop-corrected mass matrix for the neutral Higgs bosons can
be split into blocks for CP-even and CP-odd Higgs bosons as,
in this case, the mixing between them vanishes, $\hat{\Sigma}_{hA}(p^2) = \hat{\Sigma}_{HA}(p^2) = 0$.

In the Feynman diagrammatic or fixed-order approach, the self energies and
mixings 
$\hat{\Sigma}_{H_iH_j}(p^2)$ are calculated order by order in perturbation
theory by evaluating the corresponding Feynman diagrams. 
The renormalized self energies can be split into parts of different loop-order
\begin{align}
\hat \Sigma_{H_iH_j}(p^2) = \hat \Sigma^{(1)}_{H_iH_j}(p^2)
+ \hat \Sigma^{(2)}_{H_iH_j}(p^2) + \dots
\end{align}
where the superscript $(n)$, $i = 1,2,\dots$, denotes the order. At each
 order $(n)$, the renormalized self energies can be decomposed further
into an unrenormalized part and a counterterm part,
\begin{align}\label{eq:selfgen}
\hat \Sigma_{H_iH_j}^{(n)}(p^2) &= \Sigma_{H_iH_j}^{(n)}(p^2)
 +  \frac{1}{2}p^2
\left(\delta^{(n)} Z_{H_i H_j}^\dagger + \delta^{(n)} Z_{H_i H_j}\right)
\nonumber \\& \quad
 - \frac{1}{2}  \left(\left(M_{H_j}^{(0)}\right)^2 \delta^{(n)} Z_{H_i H_j}^\dagger
 + \left(M_{H_i}^{(0)}\right)^2 \delta^{(n)} Z_{H_i H_j}\right) -  \delta^{(n)} \mathcal
 M_{H_iH_j}
\nonumber \\& \quad 
 + \text{ product of counterterms, such as }\delta^{(n_1)} Z_{H_i H_j}\delta^{(n_2)} \mathcal
 M_{H_iH_j} 
\nonumber \\& \quad \quad
\text{ with } n_1 + n_2 = n
\end{align}
where the counterterm part consists of  a $Z$ factor and a mass matrix counterterm of $n$th loop
order as well as products of $Z$ factors of $n_1$th loop order and mass matrix
counterterms of $n_2$th loop order with $n_1 + n_2 = n$. (Note that counterterms of order $m<n$ also appear as insertions in $(n-m)$-loop diagrams contributing to the $n$th order unrenormalized self-energy.)

Field strength renormalization can be performed in a minimal
way by replacing the Higgs doublets by a $Z$ factor and a renormalized Higgs doublet,
\begin{align}
H_u \rightarrow \sqrt{Z_{H_u}} H_u &= \left(1 + \frac{1}{2}\delta^{[1]} Z_{H_u}
+ \frac{1}{2}\delta^{[2]} Z_{H_u}  - \frac{1}{8} (\delta^{[1]}
Z_{H_u})^2 + \dots \right) H_u \nonumber \\
&= \left(1 + \frac{1}{2}\delta^{(1)} Z_{H_u}
+ \frac{1}{2}\delta^{(2)} Z_{H_u}  + \dots \right) H_u,\\
H_d \rightarrow \sqrt{Z_{H_d}} H_d &= \left(1 + \frac{1}{2}\delta^{(1)} Z_{H_d}
+ \frac{1}{2}\delta^{(2)} Z_{H_d}  + \dots \right) H_d.
\end{align}
where the $[n]$ denotes the $n$th loop order of the expansion of the $Z$~factor
$Z_{H_q}$, $q = u, d$ with $Z_{H_q} = 1 + \frac{1}{2}\delta^{[1]} Z_{H_q}
+ \frac{1}{2}\delta^{[2]} Z_{H_q} + \dots$, while $(n)$ already takes into account the square
root with $ \delta^{(1)} Z_{H_q} = \delta^{[1]} Z_{H_q}$, $\delta^{(2)}
Z_{H_q} = \delta^{[2]} Z_{H_q}  - \frac{1}{4} (\delta^{[1]} Z_{H_q})^2$,
etc. The $Z$~factors entering  Eq.~\eqref{eq:selfgen} are obtained by the
transformation 
\begin{align}
&\delta^{(n)} Z_{H_i H_j} = \left(\mathcal U \delta^{(n)} Z_{\phi\phi} \mathcal
U^\dagger\right)_{ij}  \quad \text{with} \quad \mathcal U
= \begin{pmatrix} \mathcal U_n & 0 \\ 0 & \mathcal U_c \end{pmatrix},
 \\&
\delta^{(n)} Z^\dagger_{H_i H_j} = \left[\left(\mathcal U \delta^{(n)} Z_{\phi\phi} \mathcal
U^\dagger\right)^\dagger\right]_{ij}, 
\\
&\delta^{(n)} Z_{\phi\phi}
= \text{diag}\left(\delta^{(n)} Z_{H_d}, \delta^{(n)} Z_{H_u}, \delta^{(n)} Z_{H_d}, \delta^{(n)} Z_{H_u}, \delta^{(n)} Z_{H_d}, \delta^{(n)} Z_{H_u}\right),
\end{align}
and $\mathcal U_n$ and $\mathcal U_c$ defined in Eq.~\eqref{eq:fieldtreetrafo}.

The counterterm mass matrix $\delta^{(n)} \mathcal M_{H_iH_j} = (\mathcal U \delta^{(n)} \mathcal M_{\phi \phi}
 \mathcal U^\dagger)_{ij}$ is determined
from the mass matrices given 
in Eqs.~\eqref{eq:Mphi}--\eqref{eq:Mphizeta}\footnote{With this definition of the counterterm mass matrix, we follow the approach where the mixing angles do
not receive counterterms and can be understood as already renormalized. We could  also introduce counterterms for the mixing angles, leading to a different set of counterterms, and the explicit expressions for the $Z$ factors may be changed depending on the renormalization conditions. 
Both approaches are valid.  
}. Subsequently several approaches are possible:
\begin{itemize}
\item
We can directly introduce counterterms for the parameters appearing in 
Eqs.~\eqref{eq:Mphi}--\eqref{eq:Mphizeta},
\begin{align}\label{eq:Parorg}
m_1^2, \quad m_2^2, \quad B_\mu, \quad v_d, \quad
v_u, \quad \varphi_u,  \quad g, \quad g',
\end{align}
by the replacement
\begin{align}\label{eq:placeholder}
P \rightarrow P + \delta^{(1)} P + \delta^{(2)} P + \dots
\end{align}
where $P$ is a placeholder for the parameters appearing in
Eq.~\eqref{eq:Parorg}. Only the combinations $m_1^2
= m_{H_d}^2 + |\mu|^2$ and $m_2^2
= m_{H_u}^2 + |\mu|^2$ appear in the calculation of the Higgs boson mass
matrix. Thus $|\mu|^2$ can always be absorbed into $ m_{H_i}^2$, $i = u, d$,
and we can treat $m_1^2$ and $m_2^2$ as independent parameters, leaving  8 parameters that need to be renormalized.
For a calculation at $n$th order, counterterms up to
$\delta^{(n)} P$ have to be included. After introducing the
counterterms, the mass matrix is expanded and the part
including the counterterms can be
separated order by order into
\begin{align}
\delta \mathcal M_{\phi\phi} =  \delta^{(1)} \mathcal
M_{\phi\phi} + \delta^{(2)} \mathcal M_{\phi\phi} + \dots.
\end{align} 
where products of counterterms, such as $\delta^{(n_1)} P_1\delta^{(n_2)} P_2$
with $P_i$, $i = 1, 2$, being placeholders as in Eq.~\eqref{eq:placeholder},
are included in the mass matrix counterterm $\delta^{(n)} \mathcal
M_{\phi\phi}$, $n = n_1 + n_2$.
\item Using the parameters of Eq.~\eqref{eq:Parorg} can make comparisons with
experiment more tedious, since conversion relations between these
parameters and ones that are more easily accessible to experiment are necessary.  These relations are also affected by quantum corrections.
Instead, it can be helpful to do the conversion at the beginning of the calculation of the Higgs boson masses and to use the ``more physical'' parameters as an input. A commonly used choice of parameters is given by the weak gauge boson masses,
the ratio of the Higgs vacuum expectation values, the
electric charge $e \equiv gg'/(g^2 +g'^2)^{1/2}$, the CP-odd or the charged Higgs
boson mass, and the tadpole parameters of Eqs.~\eqref{eq:tphiu}--\eqref{eq:tzetau}. Then the 8 parameters to renormalize are:
\begin{align}
M_Z, \quad M_W, \quad e, \quad \tan \beta, \quad M_A \text{ or }  M_{H^\pm}, \quad
t_{\phi_u}, \quad t_{\phi_d}, \quad t_{\zeta_u}.
\end{align}
When converting from the original
parameters to the ones with a more physical interpretation, all parameters
should be kept, even if, at
tree-level, they are equal to another parameter or vanish,  
since the parameter relations might be changed by higher-order
contributions and vanishing
parameters might receive non-zero loop contributions. This is particularly true for the tadpole
parameters and the mixing angle $\beta_{m}$.
The conversion relations of these parameters are given as (see e.g.~\cite{Frank:2006yh}) 
\begin{align}
v_u &= \sqrt{2}\frac{M_W \sin \theta_W \sin \beta}{e}, \quad  v_d
= \sqrt{2}\frac{M_W \sin \theta_W \cos \beta}{e}, 
\\
 g &= \frac{e}{\sin \theta_W}, \quad g' = \frac{e}{\cos \theta_W},\\
B_\mu &=  \frac{\cos \beta}{\cos^2 (\beta - \beta_m)}
\Biggl \{\frac{e^2}{4
M_W^2 \sin^2 \theta_W} \frac{\cos^4(\beta- \beta_m)}{\cos^4\beta}
 t_{\zeta_u}^2 
\nonumber
\\&
+\left[
M_A^2 \sin\beta
+ \frac{e}{2 M_W \sin \theta_W}\left(\sin^2 \beta_m \tan \beta
t_{\phi_d} + \cos^2 \beta_m t_{\phi_u}\right)\right]^2 
\Biggr\}^{\frac{1}{2}},
\\
m_{H_u}^2 &= M_A^2 \frac{\cos^2 \beta}{\cos^2(\beta-\beta_m)} + \frac{1}{2}
M_Z^2 \cos(2 \beta)\\&\quad +
\frac{e \sin \beta_m}{
2 M_W \sin \theta_W \cos^2(\beta-\beta_m)} \Bigl[t_{\phi_d} 
\sin \beta_m \cos \beta
\nonumber \\& \qquad\qquad\qquad\qquad
- t_{\phi_u}\left(\sin \beta \sin \beta_m + 2 \cos \beta \cos \beta_m\right)
\Bigr],
\\  
m_{H_d}^2 &= M_A^2 \frac{\sin^2 \beta}{\cos^2(\beta-\beta_m)}
- \frac{1}{2} M_Z^2 \cos(2 \beta) 
\nonumber\\&\quad
 +
\frac{e \cos \beta_m}{2
M_W \sin \theta_W \cos^2(\beta-\beta_m)} \Bigl[t_{\phi_u}  \sin \beta \cos \beta_m
\nonumber \\& \qquad\qquad\qquad\qquad
-  t_{\phi_d} \left(\cos \beta \cos \beta_m
+ 2 \sin \beta \sin \beta_m\right)\Bigr],\\
\tan(\varphi_u) &= - \frac{t_{\zeta_u} \cos^2(\beta
-\beta_m)}{ C \cos^2 \beta}
\nonumber \\
\text{with }&
C = \frac{2 M_W \sin \theta_W \sin\beta}{e} M_A^2 +
t_{\phi_d} \tan \beta \sin^2 \beta_m
 +t_{\phi_u} \cos^2 \beta_m \;.
\end{align}
Here we have chosen the CP-odd Higgs boson mass  $M_A^2$ as an input parameter, with
$M_A^2  = (\mathcal
U_{\beta_m} \mathcal M_{\zeta} U_{\beta_m}^\dagger)_{11}$, which yields the
tree-level mass squared for tree-level parameters, i.e.\ for $t_{\phi_u} =
t_{\phi_d}=  t_{\zeta_u} = 0$ and $\beta_m = \beta$.
The weak mixing angle $\theta_W$ ($\cos \theta_W = M_W/M_Z$) has been
introduced for a compact notation. For $t_{\phi_u} =
t_{\phi_d}=  t_{\zeta_u} = 0$ and $\beta_m = \beta$, the tree-level relations
between the parameters are recovered. For the renormalization procedure, all
parameters except for the mixing angle $\beta_m$ are treated as bare
parameters and are replaced by the renormalized parameter and the
corresponding counterterm as in Eq.~\eqref{eq:placeholder}.
\end{itemize}

\begin{figure}
  \includegraphics[width=\textwidth]{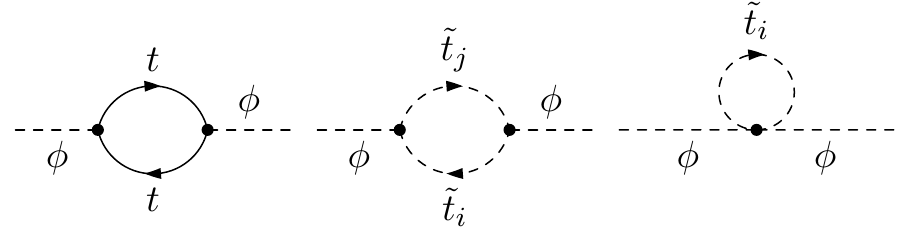}
\caption{Self-energy diagrams contributing to the $\mathcal O(\alpha_t)$,
$\alpha_x = h_x^2/(4 \pi)$, where
$\phi = h,\,H,\,A$ denoting the different Higgs bosons  and
$t$ and $\tilde t_i$ the top quarks   and top squarks, respectively.
 }\label{fig:1ltopse}
\end{figure}

The largest corrections are due to top quark/squark contributions as the
coupling between top quarks and the Higgs boson 
is proportional to the top Yukawa coupling which
is relatively large, see e.g.\
Refs.~\cite{Ellis:1990nz,Ellis:1991zd,Barbieri:1991tk,Brignole:1991pq}. Due to
the underlying supersymmetry, also the top 
squark 
coupling depends on the top Yukawa coupling. Now, let's first consider only
contributions proportional to the top Yukawa coupling squared and neglect
gauge couplings in the case of a CP-conserving MSSM, i.e. all parameters are
assumed to be real. Additionally, the CKM matrix is approximated by the unity
matrix and, hence, as real. The Feynman diagrams
 contributing to the unrenormalized self-energies are depicted in 
Fig.~\ref{fig:1ltopse}.

The counterterm mass matrix  is then
\begin{align}
&\delta \mathcal M_{H_iH_j} = (\mathcal U \delta \mathcal M_{\phi \phi}
\mathcal U^\dagger)_{ij} 
\\ \nonumber
& \text{with} \quad 
\delta \mathcal M_{\phi \phi} = 
  \begin{pmatrix} \sin^2 \beta & - \sin \beta \cos
  \beta \\ - \sin \beta \cos \beta & \cos^2 \beta \end{pmatrix} \delta^{(1)}
 {M_{A}^2} 
&\\[2ex]& \quad\ \nonumber
+ \frac{e}{2 M_W \sin {\theta}_W} \begin{pmatrix}- \cos \beta (1 + \sin^2 \beta) &- \sin^3 \beta 
\\ -\sin^3 \beta  &\cos \beta \sin^2 \beta  \end{pmatrix} \delta^{(1)} 
t_{\phi_d}
&\\[2ex]& \quad\ 
+ \frac{e}{2 M_W \sin {\theta}_W} \begin{pmatrix}  
 \cos^2 \beta \sin \beta&-  \cos^3 \beta
\\-\cos^3 \beta & - (1+ \cos^2 \beta) \sin \beta
\end{pmatrix}\delta^{(1)}  t_{\phi_u}~. \label{eq:CTmatrix1lreal}
\end{align} 
where the additional approximation of a vanishing CP-odd Higgs boson mass,
$M_A = 0$, is applied.  Without this approximation, also a $\tan \beta$
counterterm contribution proportional to $\delta^{(1)} \tan \beta$ would
appear. Conveniently, one can define $\tan \beta$ as $\overline{\text{DR}}$
parameter so that $\delta^{(1)} \tan \beta$ contains only divergent parts, and
hence, does not contribute to the finite result even if $M_A \neq 0$. 
As mentioned above, the angle $\beta_m$ does not receive a counterterm in
our approach, and in Eq.~\eqref{eq:CTmatrix1lreal} after performing the
expansion about the counterterm, $\beta_m$ has been identified with its 
tree-level value $\beta$. Similarly, the tree-level tadpole parameters have been
set to zero. In a complete one-loop calculation without the approximation of
vanishing gauge couplings, a counterterm to the Z~boson mass would also be
needed. 

We use the renormalization conditions,
\begin{enumerate}
\item that the CP-odd Higgs boson is on-shell in the approximation of $M_A =
0$,
\begin{align}
\hat{\Sigma}_{AA}^{(1)}(0) = 0,
\end{align}
which yields 
\begin{align}\label{eq:deltaMA}
\delta^{(1)} M_A^2 = \Sigma_{AA}^{(1)}(0) &= 
- \frac{3 \alpha_{\text{EM}} m_t^2}{8 M_W^2 \pi \sin^2 \theta_W \tan^2 \beta}
\nonumber\\& \quad \cdot
 \Biggl\{2 A_0(m_t^2) - 
      A_0(m_{\tilde t_1}^2)\left[1 +\frac{(A_t +\mu \tan \beta)^2}{m_{\tilde
t_1}^2 - m_{\tilde t_2}^2}\right] \nonumber\\& \qquad\qquad\quad
-     A_0(m_{\tilde t_2}^2)\left[1 -\frac{(A_t +\mu \tan \beta)^2}{m_{\tilde t_1}^2 -
      m_{\tilde t_2}^2} \right]\Biggr\},
\end{align}
where  the one-loop integral is $A_0(m^2) \equiv - 16 i \pi^2 \mu^{4-D}
 \int \frac{d^D q}{(2 \pi)^D}\frac{1}{q^2 -m^2}$, with $m$ the mass of the particle in the loop.  The dimension is denoted
 by $D = 4 - \epsilon$, and $\alpha_{\text{EM}} = e^2/(4 \pi)$. 

\item that the tadpole contributions vanish,
\begin{align}
\hat{T}^{(1)}_{\phi_i} = T^{(1)}_{\phi_i} + \delta^{(1)} t_{\phi_i} =
0 \quad \text{with} \quad i = u, d, 
\end{align}
where $\hat{T}^{(1)}_{\phi_i}$ and $T^{(1)}_{\phi_i}$ are the renormalized and
unrenormalized one-loop contributions to the one-point vertex function. These conditions result in 
\begin{align}
\delta^{(1)} t_{\phi_u} &=\frac{3 e m_t^2}{16
M_W \pi^2 \sin \beta \sin \theta_W}
\nonumber \\& \quad \cdot
\Biggl\{2 A_0(m_t^2)
- A_0(m_{\tilde t_1}^2)\left[1 + \frac{A_t (A_t - \mu \cot \beta)}{m_{\tilde
t_1}^2 - m_{\tilde t_2}^2}\right]\nonumber \\& 
\qquad \qquad \qquad \,
 - A_0(m_{\tilde t_2}^2)
       \left[1  - \frac{A_t (A_t - \mu \cot \beta)}{m_{\tilde
t_1}^2 - m_{\tilde t_2}^2}\right] \Biggr\}, \\
\delta^{(1)} t_{\phi_d} &=  \frac{3 e m_t^2 }{16
       M_W \pi^2 \sin \beta \sin \theta_W} \frac{\mu (A_t - \mu \cot \beta)}{(m_{\tilde
t_1}^2 - m_{\tilde t_2}^2)}
\nonumber \\& \quad \cdot
\left[A_0(m_{\tilde t_1}^2) - A_0(m_{\tilde t_2}^2)\right]\;.
\end{align}
\end{enumerate}
 
Applying this approximation, the self energies can be evaluated as
\begin{align}
\Sigma_{\phi_u \phi_u} &= \frac{3 \alpha_{\text{EM}} m_t^2} {8
M_W^2 \pi \sin^2 \beta \sin^2 \theta_W}\Biggl\{2(1 - D)
A_0\left(m_t^2\right) \nonumber \\& \quad + A_0\left(m_{\tilde{t}_1}^2\right) 
       \Biggl[1 +(D - 2)\frac{ m_t^2}{m_{\tilde{t}_1}^2}
       + \frac{A_t^2}{m_{\tilde{t}_1}^2 - 
          m_{\tilde{t}_2}^2}
      \nonumber   \\& \qquad\qquad
+ \frac{2(D - 2) m_t^2  A_t \left(A_t -\mu\cot \beta\right)}{m_{\tilde{t}_1}^2 (m_{\tilde{t}_1}^2 -
          m_{\tilde{t}_2}^2)} 
      \nonumber   \\& \qquad\qquad + 
        \frac{m_t^2 A_t^2 \left(A_t -\mu\cot \beta\right)^2 
          \left[(D - 6) m_{\tilde{t}_1}^2 - (D - 2) m_{\tilde{t}_2}^2\right]}{
          m_{\tilde{t}_1}^2 (m_{\tilde{t}_1}^2 - m_{\tilde{t}_2}^2)^3}\Biggr]
\nonumber \\& \quad+ 
      A_0\left(m_{\tilde{t}_2}^2\right) \Biggl[1 + 
          (D - 2) \frac{m_t^2}{m_{\tilde{t}_2}^2} 
-        \frac{A_t^2}{m_{\tilde{t}_1}^2 - m_{\tilde{t}_2}^2} 
      \nonumber   \\& \qquad\qquad\
-\frac{2  (D -
          2) m_t^2 A_t \left(A_t -\mu\cot \beta\right)}{ m_{\tilde{t}_2}^2
          (m_{\tilde{t}_1}^2 - m_{\tilde{t}_2}^2)}
\nonumber\\&\qquad\qquad
+\frac{m_t^2 A_t^2 \left(A_t -\mu\cot \beta\right)^2 
          \left[(D - 2) m_{\tilde{t}_1}^2 - (D - 6) m_{\tilde{t}_2}^2\right]}{
          m_{\tilde{t}_2}^2 (m_{\tilde{t}_1}^2 - m_{\tilde{t}_2}^2)^3}  
          \Biggr]\Biggr\},
\\
\Sigma_{\phi_u \phi_d} &= - \frac{3 \alpha_{\text{EM}} m_t^2}{ 8
          M_W^2 \pi \sin^2 \beta \sin^2 \theta_W} \nonumber \\& \quad \cdot
     \Biggl\{
 A_0\left(m_{\tilde{t}_1}^2\right) 
        \Biggl[ \frac{\mu A_t}{m_{\tilde{t}_1}^2 - m_{\tilde{t}_2}^2} 
+ \frac{(D - 2) m_t^2 \mu \left(A_t -\mu\cot \beta\right)}{
        m_{\tilde{t}_1}^2 (m_{\tilde{t}_1}^2 - m_{\tilde{t}_2}^2)}  
          \nonumber   \\&\qquad\qquad +
        \frac{m_t^2 \mu  A_t \left(A_t -\mu\cot \beta\right)^2 
          \left[(D -6) m_{\tilde{t}_1}^2 - (D - 2)
 m_{\tilde{t}_2}^2\right]}{m_{\tilde{t}_1}^2 (m_{\tilde{t}_1}^2 -
 m_{\tilde{t}_2}^2)^3}\Biggr]
\nonumber \\&\quad \,
- A_0\left(m_{\tilde{t}_2}^2\right) \Biggl[ 
         \frac{\mu A_t}{m_{\tilde{t}_1}^2 - m_{\tilde{t}_2}^2} + \frac{ (D - 2) m_t^2 \mu \left(A_t
         -\mu\cot \beta\right)}{m_{\tilde{t}_2}^2(m_{\tilde{t}_1}^2 -
         m_{\tilde{t}_2}^2)} 
\nonumber
  \\&\qquad\qquad - \frac{m_t^2\mu A_t 
           \left(A_t -\mu\cot \beta\right)^2 \left[(D - 2) m_{\tilde{t}_1}^2 - 
            (D - 6) m_{\tilde{t}_2}^2\right]}{m_{\tilde{t}_2}^2 (m_{\tilde{t}_1}^2 - m_{\tilde{t}_2}^2)^3}\Biggr] \Biggr\}    
\\
\Sigma_{\phi_d \phi_d}
          &=\frac{3 \alpha_{\text{EM}} m_t^2 \mu^2}{8 M_W^2 \pi \sin^2 \beta \sin^2 \theta_W 
     (m_{\tilde{t}_1}^2 - m_{\tilde{t}_2}^2)} 
\nonumber \\& \quad \cdot
\Biggl\{
A_0\left(m_{\tilde{t}_1}^2\right) 
       \Biggl[1 + \frac{m_t^2 \left(A_t -\mu\cot \beta\right)^2 \left[(D -6) m_{\tilde{t}_1}^2 - 
           (D - 2) m_{\tilde{t}_2}^2\right]}{m_{\tilde{t}_1}^2 
          (m_{\tilde{t}_1}^2 - m_{\tilde{t}_2}^2)^2}\Biggr]
\nonumber\\&\quad
- A_0\left(m_{\tilde{t}_2}^2\right) 
       \Biggl[1 -
\frac{m_t^2 ( A_t -\mu \cot \beta)^2 \left[(D - 2) m_{\tilde{t}_1}^2 - 
           (D - 6) m_{\tilde{t}_2}^2\right]}{m_{\tilde{t}_2}^2(m_{\tilde{t}_1}^2 - m_{\tilde{t}_2}^2)^2 
          }
\Biggr]\Biggr\}, \label{eq:selfhdhd}
\end{align}

With the approximations of vanishing gauge couplings, $M_A = 0$, 
and vanishing external momenta,\footnote{In the case of non-vanishing $M_A$, the mass
counterterm $\delta M_A^2$ would include a term proportional to the $Z$ factor of
the Higgs fields,
and the mass counterterm~\eqref{eq:CTmatrix1lreal} would include a
term proportional to $\delta^{(1)} \tan \beta$. If the $Z$ factor and $\delta^{(1)} \tan \beta$ are defined in a  $\overline{\text{DR}}$ scheme, i.e.\
they contain only divergent contributions, these contributions will
cancel the $Z$ factor dependence that enters the renormalized Higgs self
energies, so that the result will be the same as taking the approximations of vanishing gauge couplings,
$M_A = 0$, and vanishing external momenta.}
all the Higgs boson masses vanish at tree-level. Then the  renormalized
self-energies of Eq.~\eqref{eq:selfgen} simplify and can be expressed in terms
of the interaction eigenstates as
\begin{align}
\hat{\Sigma}_{\phi_i \phi_j} = \Sigma_{\phi_i \phi_j} -(\delta \mathcal
M_{\phi \phi})_{ij} \quad \text{with} \quad i,j = u,d.
\end{align}
Using Eqs.~\eqref{eq:deltaMA}--\eqref{eq:selfhdhd}, we find 
\begin{align}
\hat{\Sigma}_{\phi_u \phi_u} &= -\frac{3 \alpha_{\text{EM}} m_t^4}{2 \pi
M_W^2 \sin^2\beta \sin^2 \theta_W}
\Biggl\{\frac{1}{2}
\log\left(\frac{m_{\tilde t_1}^2 m_{\tilde t_2}^2}{m_t^4}\right)
\nonumber\\&\qquad\qquad
+ \frac{A_t^2 (A_t - \mu \cot \beta)^2}{(m_{\tilde t_1}^2 -
m_{\tilde t_2}^2)^2}\left[1 - \frac{ m_{\tilde t_1}^2 + m_{\tilde
t_2}^2}{2(m_{\tilde t_1}^2 - m_{\tilde t_2}^2)} \log\left(\frac{m_{\tilde t_1}^2}{
m_{\tilde t_2}^2}\right)\right] 
\nonumber\\&\qquad\qquad
+ \frac{A_t (A_t - \mu \cot \beta)}{m_{\tilde t_1}^2 -
m_{\tilde t_2}^2}
\log\left(\frac{m_{\tilde t_1}^2}{
m_{\tilde t_2}^2}\right)
\Biggr\}
\\
\hat{\Sigma}_{\phi_u \phi_d} &=
\frac{3 \alpha_{\text{EM}} m_t^4 \mu}{2 \pi M_W^2 \sin^2\beta \sin^2 \theta_W}
\Biggl\{\frac{(A_t - \mu \cot \beta)}{2(m_{\tilde t_1}^2 -
m_{\tilde t_2}^2)} 
\nonumber
\\&\qquad\qquad
+ \frac{A_t (A_t - \mu \cot \beta)^2}{(m_{\tilde t_1}^2 -
m_{\tilde t_2}^2)^2}\left[1 -\frac{ m_{\tilde t_1}^2 + m_{\tilde
t_2}^2}{2(m_{\tilde t_1}^2 - m_{\tilde t_2}^2)} \log\left(\frac{m_{\tilde t_1}^2}{
m_{\tilde t_2}^2}\right)\right] \Biggr\}
 \\
\hat{\Sigma}_{\phi_d \phi_d} &= 
-\frac{3 \alpha_{\text{EM}} m_t^4 \mu^2}{2 \pi M_W^2 \sin^2\beta \sin^2 \theta_W}
\nonumber \\&\quad \cdot
\frac{(A_t - \mu \cot \beta)^2}{(m_{\tilde t_1}^2 -
m_{\tilde t_2}^2)^2}\left[1 -\frac{ m_{\tilde t_1}^2 + m_{\tilde
t_2}^2}{2(m_{\tilde t_1}^2 - m_{\tilde t_2}^2)} \log\left(\frac{m_{\tilde t_1}^2}{
m_{\tilde t_2}^2}\right)\right],
\end{align}
which was derived via the effective potential approach (see
Sect.~\ref{sec:effpot}) with slightly different conventions in
Ref.~\cite{Ellis:1991zd}. 
The one-loop Higgs masses squared $(M_h^{(1)})^2$, $(M_h^{(1)})^2$ can then be
obtained by calculating the zeroes of the 
determinant of the two-point vertex function given in
Eq.~\eqref{eq:loopcorrHmassmatrix} and \eqref{eq:2ptvtxfct},
\begin{align}
\left(M_{h,H}^{(1)}\right)^2 &= \frac{1}{2}\Biggl\{\left(M_h^{(0)}\right)^2
+\left(M_H^{(0)}\right)^2 -\left(\hat{\Sigma}_{\phi_u \phi_u}
+\hat{\Sigma}_{\phi_d \phi_d}\right) \nonumber\\&\qquad
\pm \Biggl[\Bigl[\left(M_h^{(0)}\right)^2
-\left(M_H^{(0)}\right)^2 \nonumber\\&\qquad\qquad\quad
 + \cos(2\alpha)\left(\hat{\Sigma}_{\phi_d \phi_d}-\hat{\Sigma}_{\phi_u \phi_u}\right)
 + 2 \sin(2\alpha)\hat{\Sigma}_{\phi_u \phi_d}\Bigr]^2
\nonumber\\&\qquad\quad
 +
 \left[\sin(2\alpha)\left(\hat{\Sigma}_{\phi_d \phi_d}-\hat{\Sigma}_{\phi_u \phi_u}\right)
 - 2\cos(2\alpha)\hat{\Sigma}_{\phi_u \phi_d}\right]^2\Biggr]^{\frac{1}{2}}
\Biggr\}.
\end{align}
This result takes also higher-order corrections 
 into account since the expressions are not linear in the
 self energies.
 Expanding the
result for the mass of the lighter Higgs boson and 
keeping only terms of the order $\mathcal O(m_t^4)$, leads to the 1-loop
contribution to the mass squared of lightest CP-even Higgs boson, $\Delta M_h^2 =
\left(M_h^{(1)}\right)^2 - \left(M_h^{(0)}\right)^2$,
\begin{align}\label{eq:deltamhFD}
\Delta M_h^2 &= - \hat{\Sigma}_{h h} = - \left[\cos^2\alpha \hat{\Sigma}_{\phi_u \phi_u}
- \sin(2 \alpha) \hat{\Sigma}_{\phi_u \phi_d} + \sin^2\alpha \hat{\Sigma}_{\phi_d \phi_d}\right]
\nonumber
\\& = 
\frac{3 \alpha_{\text{EM}} m_t^4}{2 \pi
M_W^2 \sin^2\beta \sin^2 \theta_W}
\Biggl\{\frac{1}{2}\cos^2 \alpha \log\left(\frac{m_{\tilde t_1}^2m_{\tilde t_2}^2}{m_t^4}\right)
\nonumber
\\&\quad
- \frac{ \cos\alpha (\cos \alpha A_t + \mu \sin \alpha) (A_t - \mu \cot \beta)}{m_{\tilde t_1}^2 -
m_{\tilde t_2}^2}\log\left(\frac{m_{\tilde t_1}^2}{
m_{\tilde t_2}^2}\right)
\nonumber
\\&\quad
+ 
\frac{(A_t - \mu \cot \beta)^2(\cos \alpha A_t + \mu \sin \alpha)^2}{(m_{\tilde t_1}^2 -
m_{\tilde t_2}^2)^2}
\nonumber \\&\qquad \cdot
\left[1 -\frac{ m_{\tilde t_1}^2 + m_{\tilde
t_2}^2}{2(m_{\tilde t_1}^2 - m_{\tilde t_2}^2)} \log\left(\frac{m_{\tilde t_1}^2}{
m_{\tilde t_2}^2}\right)\right]
\Biggr\}.
\end{align}  

\subsection{The Effective Potential Approach}\label{sec:effpot}

The effective potential offers another method for calculating the fixed-order self energies of the Higgs
bosons. The second derivatives of $V_{\text{eff}}$ with respect to the external fields gives the corresponding self energies in the limit of vanishing external
momenta. While fewer diagrams have to be
calculated, the field dependence of masses and couplings has to be preserved throughout (and to include the momentum dependence, the approach of the previous section must be used.) Typically, the effective potential is computed in dimensional
reduction, leading to masses in the $\overline{\text{DR}}$ scheme. A conversion of schemes can be performed to yield results in an on-shell scheme.

\begin{figure}
\begin{center}
\begin{tabular}{ccccccccc}
\raisebox{0.6cm}[-0.6cm]{$\bullet$}  & \raisebox{0.6cm}[-0.6cm]{$+$} &
  \includegraphics[width=0.12\textwidth]{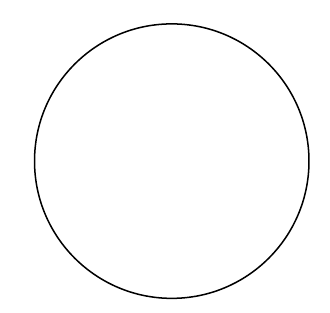}& 
\raisebox{0.6cm}[-0.6cm]{$+$} &
 \includegraphics[width=0.12\textwidth]{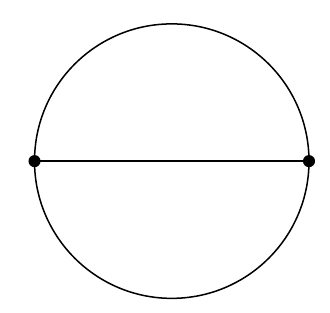} & 
\raisebox{0.6cm}[-0.6cm]{$+$} & 
\includegraphics[width=0.24\textwidth]{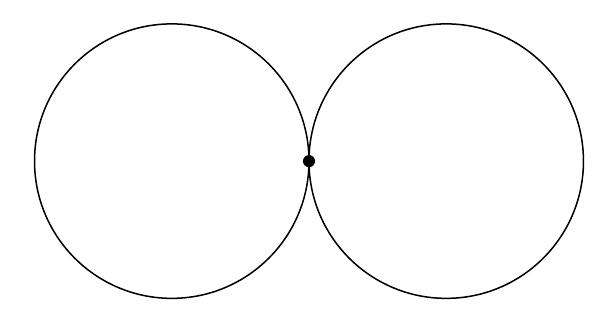} & 
\raisebox{0.6cm}[-0.6cm]{$+$} &\raisebox{0.6cm}[-0.6cm]{ \dots} \\[1ex]
(a) & & (b) & & (c) & & (d) & &
\end{tabular}
\end{center}
\caption{Diagrammatic representation of the effective potential.
 }\label{fig:effpot}
\end{figure}

The effective potential can be expressed as a sum of one-particle irreducible 
Green functions $\Gamma^{(n)}$ at vanishing external momenta,
\begin{align}
V_{\text{eff}}(\psi_{\text{cl}}) =
- \sum_{n=0}^\infty \frac{1}{n!} \psi_{\text{cl}}^n \Gamma^{(n)}(p_i = 0),
\end{align}
where $\psi_{\text{cl}}$ denotes the classical fields. A derivation of this
expression can be found in the review~\cite{Quiros:1999jp}. The effective
potential 
can then be depicted by  vacuum diagrams as in
Fig.~\ref{fig:effpot} where the dot (a) is the tree-level potential and the
next diagrams are the one-loop (b), two-loop (c) and (d), \dots contribution. 
Each diagram represents the sum over all diagrams with any 
possible number of external Higgs-boson legs with vanishing momenta.

Performing the sum over all the one-loop diagrams leads to the
Coleman--Weinberg potential~\cite{Coleman:1973jx},
\begin{align}\label{eq:1lColWeinpot}
V^{(1)} = \frac{1}{64 \pi^2} \sum_n (-1)^{2
s_n} x_n \left(m_n^2\right)^2 \left[\ln \left(\frac{m_n^2}{Q^2}\right) - c_n\right], 
\end{align}
where $s_n = 0, 1/2, 1$ for scalar, fermion and gauge fields, the $m_n$ are the
corresponding field-dependent masses, the $x_n$ are the number of degrees of freedom (spin, charge, color, etc.), and $Q$  is the renormalization scale. The
constant $c_n$ depends on the 
 regularization and renormalization scheme, see e.g.~\cite{Martin:2001vx}. In
 the $\overline{\text{DR}}$ scheme the constant is $3/2$.\footnote{Or to be more precise, the
 $\overline{\text{DR}}'$ scheme. The difference between
 the $\overline{\text{DR}}$ and the $\overline{\text{DR}}'$ scheme is the
 treatment of the epsilon scalars which in the $\overline{\text{DR}}$ scheme
 have to be taken explicitly into account while in the $\overline{\text{DR}}'$
 their effect is implicitly taken care of by using redefined soft-breaking
 parameters. The pure fermion and sfermion loops are the same in both schemes,
 so, focusing on the top/stop contribution at one-loop, we do not need to
 distinguish between these schemes.}
 
Very specific to the one-loop effective potential is the appearance of the
logarithm, due to the cyclic symmetry of the diagram (b) in
Fig.~\ref{fig:effpot} which leads to a factor of $1/j$ when $j$ couplings with
external Higgs-boson legs,  $j =1,\dots, \infty$, are inserted. The effect of these external couplings is
 absorbed into the field-dependent masses in Eq.~\eqref{eq:1lColWeinpot}. At higher orders, as depicted
in diagrams (c) and (d) in Fig.~\ref{fig:effpot}, the propagators are always
attached to an internal vertex and the insertion of $j$ couplings with
external legs to a propagator will lead to a geometric series which can be
rewritten in terms of a propagtor with field-dependent mass, see e.g. the
appendix of Ref.~\cite{Coleman:1973jx}. Thus, for the
calculation of higher-order corrections, the complete contribution at a
given order is obtained by calculating all vacuum diagrams at that order and replacing
the masses and couplings by their 
field-dependent counterparts. 

The path integral offers a complementary approach to deriving the effective potential, as shown in Ref.~\cite{Jackiw:1974cv} and reviewed in Ref.~\cite{Quiros:1999jp}.

We can make contact with the Feynman-diagrammatic calculation by computing the one-loop
contributions 
at $\mathcal O(m_t^4)$ in the effective
potential approach. The top/stop contributions to the effective potential can be obtained by using
Eq.~\eqref{eq:1lColWeinpot},
\begin{align}
V^{(1)} =V^{(1)}_t + V^{(1)}_{\tilde t}\;,
\end{align}
with
\begin{align}
V^{(1)}_t &= - \frac{4N_c}{64 \pi^2} \left[m_t(H_q)\right]^4 \left[\ln \left(\frac{\left[m_t(H_q)\right]^2}{Q^2}\right)
- \frac{3}{2}\right], \\
V^{(1)}_{\tilde t} &=  \frac{2N_c}{64 \pi^2} \sum_{j=1}^2  \left[m_{\tilde
t_j}(H_q)\right]^4 \left[\ln \left(\frac{\left[m_{\tilde t_j}(H_q)\right]^2}{Q^2}\right) - \frac{3}{2}\right] 
\end{align}
where $q = u, d$, the color factor $N_c = 3$, and $m_t(H_q)$ and $m_{\tilde t_j}(H_q)$ denote the
field-dependent top and stop masses, respectively.
These field-dependent masses
are given in the gaugeless 
limit as
\begin{align}
m_t(H_q) = h_t  |H_u^0|, \quad H_u^0 =
\text{e}^{\text{i} \varphi_u} \left[v_u +
  \frac{1}{\sqrt{2}}({\phi}_u + i {\zeta}_u)\right]
\end{align}
and
\begin{align}
&\mathcal M_{\tilde t}(H_q) = 
\begin{pmatrix}
m_{{\tilde Q}_3}^2 + h_t^2  |H_u^0
|^2 &
h_t \left[A_t^* 
\left(H_u^0\right)^*- 
\mu H_d^0 \right]\\
h_t \left[A_t H_u^0 - \mu^* \left(H_d^0\right)^* \right]
&
m_{{\tilde u}_3}^2 + h_t^2  |H_u^0|^2 
\end{pmatrix}\\
&\text{with } H_d^0  = v_d + \frac{1}{\sqrt{2}}({\phi}_d - i{\zeta}_d),
\end{align}
respectively. The expressions of $H_u^0$ and $H_d^0$ are the neutral
components of Eq.~\eqref{eq:Higgsdoublets}. Restricting to the
CP-conserving case,
the field dependent stop masses squared are
\begin{align}
\left[m_{\tilde t_j}(H_q)\right]^2 &= \frac{1}{2}\left(m_{{\tilde Q}_3}^2 + m_{{\tilde u}_3}^2\right) + h_t^2
\left[\left( v_u +  \frac{1}{\sqrt{2}}{\phi}_u\right)^2 + \left(\frac{1}{\sqrt{2}}\zeta_u\right)^2\right] \nonumber \\& \pm
\sqrt{\left(m_{{\tilde Q}_3}^2 - m_{{\tilde u}_3}^2\right)^2 + h_t^2
\left|A_t \left[ v_u +\frac{1}{\sqrt{2}}\left(
  {\phi}_u + i {\zeta}_u\right)\right] - \mu^* \left[v_d +\frac{1}{\sqrt{2}}\left( 
  {\phi}_d + i {\zeta}_d\right)\right] \right|^2\;.
}
\end{align}
The different one-loop contributions can then be obtained by taking field derivatives. The tadpole contributions are 
\begin{align}
T_{\Phi_i}^{(1)} = - \frac{\partial V^{(1)}}{\partial \Phi_i} \quad \text{with} \quad \Phi_i = \phi_u, \phi_d, \zeta_u, \zeta_d
\label{eq:Vefftadpole}
\end{align} 
and the self energies (with vanishing external momentum) are
\begin{align}
\Sigma_{\Phi_i \Phi_j}(0) = - \frac{\partial^2 V^{(1)}}{\partial \Phi_i\partial \Phi_j}.
\label{eq:Veffsigma}
\end{align}
It is not difficult to check that Eqs.~(\ref{eq:Vefftadpole}) and~(\ref{eq:Veffsigma}) reproduce the tadpoles and self-energies found in the previous section.

\subsection{State of the Art
}

\stepcounter{footnote}
\begin{table}[h]
\begin{tabular}{|l||c|c|c|c||c|c|c|c|}
\hline 
&\multicolumn{4}{c||}{real parameters} & \multicolumn{2
}{c|}{complex parameters}\\
&\multicolumn{2}{c|}{$\overline{\text{DR}}$~scheme} &  \multicolumn{2}{c||}{OS/mixed schemes} &
\multicolumn{2}{c|}{OS/mixed
schemes}\\ 
&$p^2 = 0$ & $p^2 \neq 0$ &
$p^2 = 0$ & $p^2 \neq 0$ &
$p^2 = 0$ & $p^2 \neq 0$ \\\hline
{ one-loop} & { } & { } & &  \cite{Chankowski:1991md,Chankowski:1992er,Dabelstein:1994hb} & { } 
& \cite{Frank:2006yh}\\\hline 
{ two-loop} &    &    &    & &    &    
\\
{ $\mathcal O (\alpha_t \alpha_s)$} 
 & \cite{Zhang:1998bm,Espinosa:1999zm,Degrassi:2001yf}  & \cite{Degrassi:2014pfa}
 & \cite{Heinemeyer:1998np,Espinosa:1999zm, Degrassi:2001yf}  & \cite{Borowka:2014wla, Degrassi:2014pfa, Borowka:2015ura}& 
 \cite{Heinemeyer:2007aq}  & { }
 \\
{ $\mathcal O (\alpha_t^2)$} 
 &  \cite{Espinosa:2000df} & { } & \cite{Espinosa:2000df,Brignole:2001jy}  & { }&  
\cite{Hollik:2014wea, Hollik:2014bua} & { }
 \\
{ $\mathcal O (\alpha_b \alpha_s)$} 
 & { } & { } & \cite{Brignole:2002bz,Heinemeyer:2004xw}  & { }& { } & { } 
 \\
{ $\mathcal O (\alpha_t \alpha_b, \alpha_b^2, \alpha_\tau^2)$} 
 & { } & { } & \cite{Dedes:2003km} & { } & { } & { } 
 \\
{ $\mathcal O (\alpha_\tau \alpha_b)$} 
 & { } & { } &  \cite{Allanach:2004rh} & { } & { } & 
\\
{ full }
 & \cite{Martin:2002wn}
&  &  & & { } & { } 
\\
{ first five rows + $\mathcal O (\alpha_{\text{EW}} \alpha_s )$}
 &  & \cite{Martin:2004kr} &  & & { } & { } 
\\\hline 
{ three-loop} &    &    &    & &    &      
\\ 
{ $O(\alpha_t \alpha_s^2)$}
&  & { } & \cite{Harlander:2008ju, Kant:2010tf} & { }& { } & { }  
\\ 
{ $O(\alpha_t \alpha_s^2, \alpha_t^2 \alpha_s, \alpha_t^3)|_{\text{LL,NLL}}$}
& \cite{Martin:2007pg} & { } &  & { }& { } & { }  
\\\hline 
\end{tabular}
\caption{\label{tab:FDcontr} Known fixed-order contributions to the
Higgs boson mass spectrum.  The results of Ref.~\cite{Martin:2002wn} and \cite{Martin:2004kr} can in principle be taken over to complex parameters; however, they have not been
analyzed for this case. $\alpha{_\text{EW}}$ denotes terms of order $g^2$ or $g^{\prime 2}$.}
\end{table}

Much effort has gone into the precise calculation of the MSSM Higgs masses in the Feynman diagrammatic approach.
In Table~\ref{tab:FDcontr}, the different known contributions are summarized. 
The one-loop corrections are completely known, including full momentum
dependence, for real and complex parameters~\cite{Chankowski:1991md,Chankowski:1992er,Dabelstein:1994hb,Frank:2006yh} and in different renormalization schemes. The one-loop corrections can be very large, up to several tens of GeV, even for superpartner masses of order 1 TeV. As discussed above, the dominant contributions come from the top quarks and squarks (which do not enter the Higgs sector at the tree level, and  therefore are of
leading order at the one-loop level.) Except for the gluons and the gluinos, all 
particles contribute to at least one of the Higgs boson masses at one-loop level, while the gluons and gluinos
appear only at two-loop level. 
The main contributions at the two-loop level are 
$\mathcal O(\alpha_t \alpha_s)$ 
and can be obtained in the approximation of vanishing gauge couplings and vanishing external momenta, which, to be more precise, amounts to a contribution to $m_h$ of  $\mathcal O(m_t^2 \alpha_t \alpha_s)$.
These corrections can also be of several GeV. They have been calculated within
a $\overline{\text{DR}}$ scheme \cite{Zhang:1998bm,Espinosa:1999zm,Degrassi:2001yf} as well as
in the 
on-shell scheme \cite{Heinemeyer:1998np, Espinosa:1999zm,Degrassi:2001yf}. Switching between
the different renormalization schemes provides an estimate of the theoretical
uncertainty from missing higher-order corrections. However, care has to be taken that the renormalization scheme
provides a good expansion point, where no corrections are
so large that convergence is spoiled.
The corrections of $\mathcal O(m_t^2 \alpha_t \alpha_s)$ are also known for
the CP-violating MSSM~\cite{Heinemeyer:2007aq}. These contributions  not
only change the values of the Higgs masses but can also modify the CP-mixing
of the Higgs boson. Recently, the calculation of the terms of $\mathcal
O(\alpha_t \alpha_s)$ without the approximation of vanishing external momenta
became available with a pure
$\overline{\text{DR}}$ renormalization scheme \cite{Degrassi:2014pfa} 
and a mixed scheme~\cite{Borowka:2014wla,
Degrassi:2014pfa,Borowka:2015ura}. Within these contributions, also terms of
order $\mathcal O(M_A^2 \alpha_t \alpha_s)$ appear. Additionally, in
Ref.~\cite{Degrassi:2014pfa} the approximation of vanishing gauge couplings is
dropped. The corrections are usually of the order of a couple of 100 MeV,
small in comparison to the contributions of $\mathcal
O(m_t^2 \alpha_t \alpha_s)$ but still relevant with respect to the already-reached 
experimental accuracy in $m_h$.  

Another important contribution at the two-loop level appears at $\mathcal
O(\alpha_t^2)$, or to be more precise, $\mathcal
O(m_t^2 \alpha_t^2)$. These contributions have been calculated in the gaugeless, $p^2=0$ limit. They are
known for both real~\cite{Espinosa:2000df,Brignole:2001jy} and complex~\cite{Hollik:2014wea, Hollik:2014bua} parameters. The corrections can be sizable, of order a
few GeV. The phase dependence in the complex case becomes particularly important for
large values of $\mu$.

For large $\tan \beta$, contributions from bottom quarks and squarks can become
sizable. The dominant correction is $\mathcal
O(\alpha_b \alpha_s)$~\cite{Brignole:2002bz,Heinemeyer:2004xw}. A good
approximation for these contributions at one-loop level can be obtained by
using a 
$\overline{\text{DR}}$ bottom quark mass with $\Delta_b$ contributions
are resummed~\cite{Carena:1999py}. In a large part of parameter space, this
approximation lies within the uncertainty band which can be obtained by
using different renormalization schemes for the bottom quark/squark sector~\cite{ Heinemeyer:2004xw}. 

Further two-loop corrections include $\mathcal O (\alpha_t \alpha_b, \alpha_b^2, \alpha_\tau^2)$~\cite{Dedes:2003km} and $\mathcal O(\alpha_\tau \alpha_b)$~\cite{Allanach:2004rh}, calculated in the gaugeless, $p^2=0$ limit, and are available in different renormalization schemes. These contributions are again most important for large $\tan \beta$ and can account for a couple of GeV. 

In the $\overline{\text{DR}}$ scheme, the complete two-loop contribution
with vanishing external momenta was given in~\cite{Martin:2002wn}. This
calculation was performed with the effective potential\footnote{The vev parameters in~\cite{Martin:2002wn} minimize the full two-loop potential, while in a number of other calculations (including, for example,~\cite{Espinosa:2000df,Degrassi:2014pfa}, and the one-loop calculation above), the vevs minimize the tree-level potential and tadpole terms are taken into account in the computation of the spectrum. Both approaches are valid; however, the meaning of the parameters is different in the two cases, which has to be accounted for when comparing or combining the results.}, and
the corrections going beyond the Yukawa corrections improve the
renormalization scale dependence of the resulting Higgs mass. However, purely electroweak terms of order $g^4$, $g'^2g^2$, $g'^4$ should be of similar order to omitted terms at finite external momentum.
Momentum-dependent effects have also been computed
at two-loop level for the terms of order $\mathcal O(\alpha_t \alpha_s)$ through $\mathcal O(\alpha_\tau \alpha_b)$ in the table above plus terms of order $\mathcal O(\alpha_s \alpha_{EW})$
\cite{Martin:2004kr}. The momentum effects have been found to be of
order a few 100 MeV in typical examples.

\begin{figure}[t]
\includegraphics[width=\textwidth]{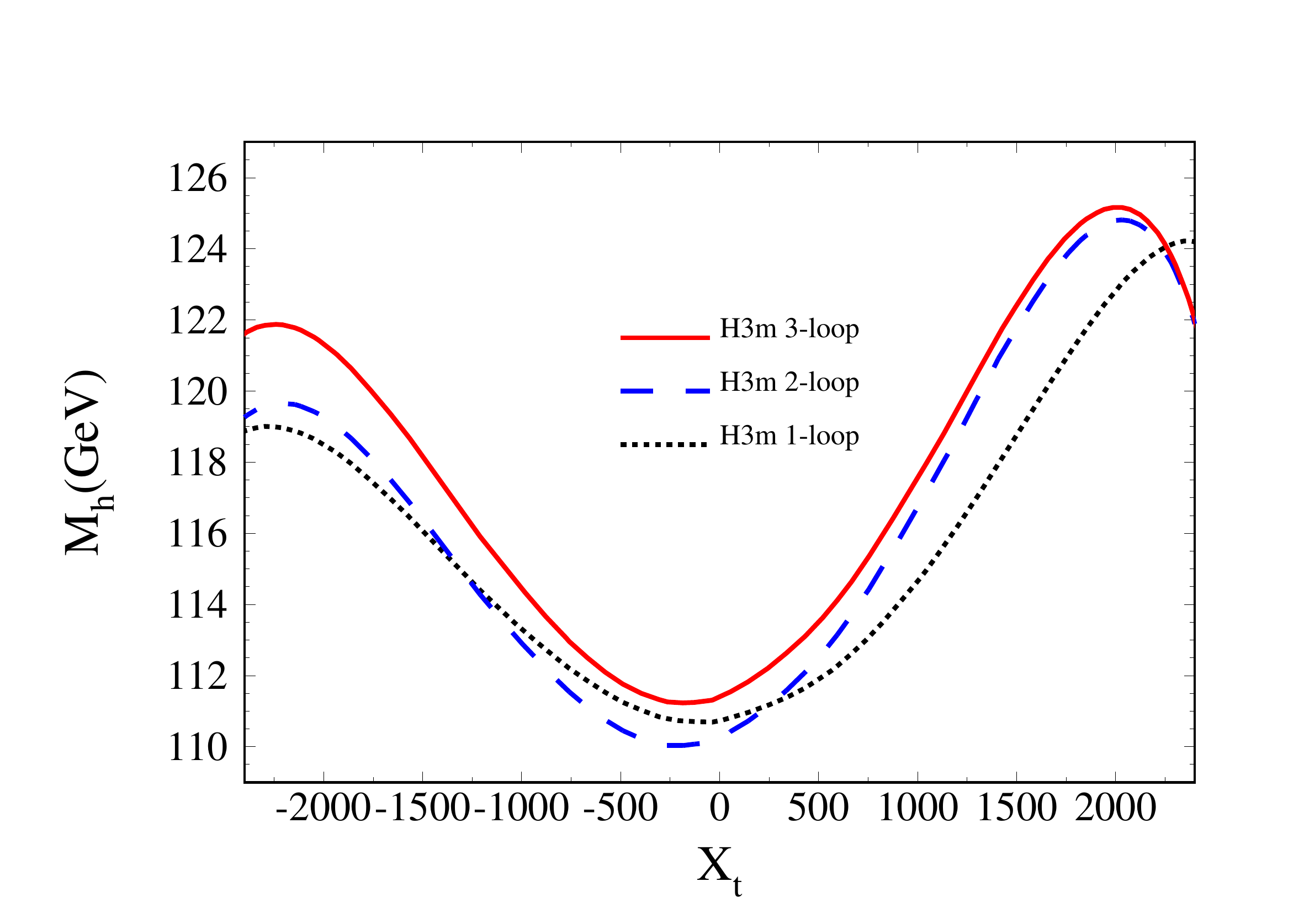}
\caption{\label{fig:fixedorderXt} The mass of the lightest Higgs boson as a function of
the squark mixing parameter $X_t = A_t - \mu \cot \beta$, for three fixed-order computations: the one-loop
 mass, the mass including  two-loop
corrections of order $\mathcal
O(\alpha_t \alpha_s,  \alpha_t^2, \alpha_b \alpha_s, \alpha_t \alpha_b, \alpha_b^2)$
 and the three-loop mass of order $\mathcal O(\alpha_s^2 \alpha_t)$. The
 renormalization scale is  
$Q = m_t$. The
fixed parameters are $M_{\text{SUSY}} = m_{{\tilde Q}_3} =  m_{{\tilde
u}_3} = 1$~TeV, $M_2 = 200$~GeV, $M_3 = 1.5$~TeV, $M_A = 500$~GeV, $\tan \beta = 10$, and $\mu = 1$~TeV.
The top quark mass is set to 173.2~GeV.
 This plot was made using the program H3m~\cite{Kant:2010tf} and
taken from Ref.~\cite{Mihaila:2015}. }
\end{figure}

At three-loop order,  including non-logarithmic terms, only contributions of order $\mathcal
O(\alpha_t \alpha_s^2)$ ($\mathcal
O(m_t^2 \alpha_t \alpha_s^2)$) are known~\cite{Harlander:2008ju, Kant:2010tf}. These terms are calculated for
vanishing external momenta and gauge couplings. For stop masses of 1~TeV, the
corrections can amount to more than  1~GeV, as shown in Fig.~\ref{fig:fixedorderXt}, and
the scale uncertainty is significantly 
reduced. Beyond this order, only logarithmic terms are
known (as discussed in Sec.~\ref{sec:RGE}). In Ref.~\cite{Martin:2007pg}, for example, the
leading and next-to leading logarithms of the order $\mathcal
O(\alpha_t \alpha_s^2, \alpha_t^2 \alpha_s, \alpha_t^3)$ and similar
four-loop order terms have been evaluated. Between these different 
contributions, cancellation effects have been observed at both three- and four-loop order~\cite{Martin:2007pg,Draper:2013oza}, leading to an overall
effect of a few hundred MeV for stop masses in the range of 1~TeV.

In Fig.~\ref{fig:fixedorderXt} the lightest Higgs mass at three loops is shown
as a function of the stop mixing parameter including the three-loop terms of
$O(m_t^2 \alpha_t \alpha_s^2)$, and compared to the one-loop level and
two-loop values. The mass values have been generated by the program
H3m~\cite{Kant:2010tf} and 
taken from Ref.~\cite{Mihaila:2015}. The corrections going beyond the orders
$\mathcal O(\alpha_t)$, $\mathcal O(\alpha_t \alpha_s)$ and $\mathcal
O(\alpha_t^2)$, at one loop in particular contributions from the electroweak
and at one- and two-loop contributions from the
bottom/sbottom sector, are taken over from 
FeynHiggs~\cite{Heinemeyer:1998yj,Heinemeyer:1998np,Degrassi:2002fi,Frank:2006yh,Hahn:2013ria}. It should be
 noted that the relative size and
signs of the 
corrections at different loop levels depend on the 
renormalization scale and scheme, taken to be $Q = m_t$ in the $\overline{\text{DR}}$ scheme in Fig.~\ref{fig:fixedorderXt}.

\section{Radiative Corrections: Renormalization Group Equation Approach}
\label{sec:RGE}

\subsection{Introduction}
\quad When supersymmetry is much more strongly broken than electroweak
symmetry, e.g., $m_{\tilde{t}}\gg m_t$, the Higgs mass may be efficiently and
accurately calculated with effective field theory techniques.  The appropriate
sequence of nonsupersymmetric EFTs descending from the UV supersymmetric
theory is determined by the hierarchies in the spectrum. For example, in the
simplest so-called ``Heavy SUSY" limit of the MSSM, all of the soft
SUSY-breaking masses as well as the CP-odd scalar mass $m_A$ lie around some
characteristic scale $m_S\gg m_t$, and the EFT valid at lower energies is just
the Standard Model with its ordinary single Higgs doublet. Because $m_A$ is
large in this scenario, we can refer unambiguously to ``the" Higgs, meaning
the light state $h$ in the SM~EFT. Another example is that of a ``split"
spectrum, where the soft scalar masses are much larger than the soft gaugino
masses, which may or may not be much larger than $m_t$. Such spectra are
motivated theoretically by focus-point
scenarios~\cite{Feng:1999mn,Feng:1999zg} and anomaly
mediation~\cite{Giudice:1998xp,Randall:1998uk} as well as
phenomenologically~\cite{Wells:2004di,ArkaniHamed:2004fb}. For a third
example, either of these spectra may be modified by taking $m_A$ of order
$m_t$, so that the theory around $m_t$ is a Two Higgs Doublet Model (2HDM), with or without extra fermions.

\quad EFT repackages the bulk of the computation of the Higgs mass into the solution of a family of renormalization group equations (RGEs). The primary virtue of this reorganization is that we can capture radiative corrections of the form $\alpha_i^{n+m-1}\log^{n}(m_S/m_t)$ for all $n$ by using $m$-loop beta functions in the RGEs for the couplings $\alpha_i$.  When there is a substantial hierarchy, the logarithms are large, dominating the quantum effects. Early computations of the radiative corrections to $m_h$ utilizing EFT and RG techniques include~\cite{Casas:1994us,Carena:1995bx,Carena:1995wu,Haber:1996fp,Carena:2000dp}, and up-to-date computations incorporating most known effects were performed in~\cite{Martin:2007pg,Giudice:2011cg,Degrassi:2012ry,Hahn:2013ria,Draper:2013oza,Bagnaschi:2014rsa,Vega:2015fna}.

\quad In this section we use effective field theory and renormalization group techniques to calculate the leading radiative corrections to the SM-like Higgs mass (henceforth, ``the Higgs mass") in the MSSM~in the example of Heavy SUSY.  We explain first the simplest estimate that gives the most important contributions to $m_h$, then discuss the many improvements that can be implemented to capture subleading but quantitatively important effects.

\subsection{Simplest Estimate}
\quad We first estimate the Higgs mass in Heavy SUSY~with a calculation that captures only the tree-level effects and the leading logarithmic quantum corrections proportional to the top Yukawa and the QCD gauge coupling. In doing so, we will be entitled to ignore many subleading but interesting considerations related to threshold corrections and renormalization scheme dependence. In later subsections we return to these issues.

\quad We begin by integrating out all of the MSSM~degrees of freedom at the renormalization scale $Q=m_S$ and matching on the SM~at tree level. The normalization of the parameters in the SM Higgs potential is convention dependent; here we use
\begin{align}
V(\Phi)=-\frac{m^2}{2}|\Phi |^2+\frac{\lambda}{2}|\Phi |^4\;,\;\;\;\Phi=\binom{0}{v+\frac{h}{\sqrt2}}\;, \;\;\; v \simeq 174 {\rm~GeV}.
\end{align}
The SM~Higgs quartic coupling $\lambda$ and the SM~Yukawa coupling $y_t$ are given at the scale $m_S$ by
\begin{align}
\label{eq:treematch}
\lambda=\frac{1}{4}(g^2+g^{\prime 2})c^2_{2\beta}\;, \;\;\;\;y_t=h_ts_\beta\;.
\end{align}
The simple form of the quartic is the result of the decoupling limit $m_A\gg m_Z$. Here $g$ and $g^\prime$ are the electroweak gauge couplings at $m_S$, which at tree level are the same in the two theories (as is the strong coupling $g_3$), and $h_t$ is the running top Yukawa at $m_S$ in the MSSM. 

\quad After matching, the SM~couplings can be run down to the next threshold, $m_t$, using the coupled set of RGEs. If we compute the tree-level Higgs mass in the SM~at this scale, $m_h=2\lambda(m_t) v^2$, 
we will have effectively captured leading logarithmic radiative corrections to $m_h$ proportional to $\log(m_S/m_t)$ that appear in the fixed-order calculation at each order in perturbation theory. These corrections are absorbed into the running coupling $\lambda$ by the RGEs. 

\quad Keeping only terms proportional to $\lambda$, $y_t$, and $g_3$, the relevant 1-loop beta functions in the SM~are:
\begin{align}
\beta^{(1)}_\lambda=12\lambda^2+12\lambda y_t^2-12 y_t^4\;,\;\;\;\;\beta^{(1)}_{g_3}=-7 g_3^3\;,\;\;\;\;\beta^{(1)}_{y_t}= y_t\left(\frac{9}{2}y_t^2-8g_3^2\right)\;,
\label{eq:simpleRGEs}
\end{align}
where the RGE for a generic coupling $y$ is
\begin{align}
\label{eq:RGEs}
\frac{dy}{dt}=\beta_y\;,\;\;\;t\equiv\log(Q)\;,\;\;\;\;\beta_y=\kappa \beta_y^{(1)}+\kappa^2\beta_y^{(2)}+\dots\;,
\end{align}
and $\kappa $ is the loop-counting factor $(16\pi^2)^{-1}$.
At this stage, we can either solve the RGEs numerically or analytically to obtain the couplings $y(t)$ as a function of boundary conditions $y(\ttil)$ at the scale $\ttil\equiv\log(m_S)$. For complicated sets of RGEs, a general analytic solution is typically not possible, but a numerical solution will give the Higgs mass with the highest precision in the EFT approach.

The RGEs can also be used to derive approximate perturbative solutions in powers of $t-\ttil$. Although the numerical solution is the most precise, the approximate analytic results are useful, because they provide the leading analytic terms when logs are large and they provide intuition for the roles of different parameters in determining the Higgs mass.  With 1-loop (2-loop, 3-loop, ...) beta functions, the perturbative formulae capture the leading logarithmic (next-to-leading logarithmic, next-to-next-to-leading logarithmic, ...) terms of a fixed-order calculation within the EFT.\footnote{With appropriate matching of parameters, the EFT fixed-order calculation can be compared with the Feynman-diagrammatic computation in the full MSSM.}

\quad We can illustrate the numerical and analytical approaches using the strong and top sector RGEs in Eq.~(\ref{eq:simpleRGEs}). To organize the analytic calculation, it is convenient to write the full beta function for each coupling $y$ evaluated at a scale $t$ both as a power series in loops, as in Eq.~(\ref{eq:RGEs}), and as a Taylor series about $\ttil$:
\begin{align}
\beta_{y}(t)=\sum_{n=1}^\infty\kappa^n\sum_{k=0}^\infty\frac{\beta_y^{(n,k)}(\ttil)}{k!}(t-\ttil)^k\;,
\end{align}
where
\begin{align}
\beta_y^{(n,k)}(t)\equiv\frac{d^k\beta_y^{(n)}}{dt^k}(t)\;.
\end{align}
We denote $\beta_y^{(n,0)}\equiv\beta_y^{(n)}$ for short. Note that $t$-derivatives of beta functions replace factors of couplings with factors of beta functions themselves, and therefore $\beta_y^{(n,k)}$ begins at order $\kappa^k$ in the loop expansion. With this decomposition, the power series solution for $y(t)$ is
\begin{align}
y(t)=y(\ttil)-\sum_{n=1}^\infty\kappa^n\sum_{k=0}^\infty(-1)^k\frac{\beta_y^{(n,k)}(\ttil)}{(k+1)!}L^{k+1}\;,
\end{align}
where $L\equiv(\ttil-t)$. Truncating to 1-loop beta functions,
\begin{align}
y(t)=y(\ttil)-\kappa\sum_{k=0}^\infty(-1)^k\frac{\beta_y^{(1,k)}(\ttil)}{(k+1)!}L^{k+1}\;,
\end{align}
and $\beta_y^{(1,k)}\propto \kappa^k$, so that the $m^{th}$ term in the series is of order $(\kappa L)^m$. With this expression and the beta functions~(\ref{eq:simpleRGEs}), it is straightforward to write a leading-log perturbative solution for $\lambda(t)$:
\begin{align}
\lambda(t)=\lambda(\ttil)-\kappa\beta_\lambda^{(1)}(\ttil)L+\kappa\frac{\beta_\lambda^{(1,1)}}{2!}(\ttil)L^2-\kappa\frac{\beta_\lambda^{(1,2)}}{3!}(\ttil)L^3+\dots\;.
\label{eq:lambdall}
\end{align}
Derivatives of the beta functions may be evaluated using the chain rule and Eq.~(\ref{eq:RGEs}). For example,
\begin{align}
\beta_\lambda^{(1,1)}=\kappa\left[12\beta_\lambda^{(1)}(2\lambda+y_t^2)+24\beta_{y_t}^{(1)}y_t(\lambda-2y_t^2)\right]+\mathcal{O}(\kappa^2)\;.
\end{align}
Fixing $t=\log(m_t)$ and $\ttil=\log{m_S}$, we can get an explicit formula for $\lambda(m_t)$ in terms of $\log(m_S/m_t)$ and the SM~couplings at $m_S$. At 2-loop order, retaining only the leading $y_t$- and $g_3$-dependent terms, we find
\begin{align}
\label{eq:simplelambdamt}
\lambda(m_t)=\lambda(m_S)+ 12 \kappa \left(y_t^4-\lambda y_t^2\right) L+\kappa^2\left(-180 y_t^6+192g_3^2y_t^4\right)L^2\;.
\end{align}
In Eq.~(\ref{eq:simplelambdamt}), the couplings in the 1-loop radiative terms on the right-hand side are to be evaluated at $m_S$, and $L\equiv\log(m_S/m_t)$.  

\quad To convert $\lambda(m_t)$ to a running Higgs mass $m_h^2(m_t)$ (which differs from the pole mass by corrections discussed in the next section), we have only to multiply by $2v^2\approx 246$~GeV$^2$:
\begin{align}
\label{eq:simplemhmt}
m_h^2(m_t)=m_Z^2 c^2_{2\beta}+ 24 \kappa \left(\frac{m_t^4}{v^2}-\frac{m_t^2m_Z^2}{2v^2}c^2_{2\beta}\right) L+24\kappa^2\frac{m_t^4}{v^2}\left(-15 y_t^2+16g_3^2\right)L^2\;,
\end{align}
 where again the parameters on the right-hand side are to be evaluated at $m_S$, and the tree-level term is obtained from Eq.~(\ref{eq:mh02}) in the limit $m_A\gg m_Z$.
 
 \quad In Eq.~(\ref{eq:simplemhmt}) we recognize the famous 1-loop log-enhanced term $m_t^4\log(m_S/m_t)$, previously obtained with the diagrammatic method in Eq.~\ref{eq:deltamhFD} and originally calculated in~\cite{Haber:1990aw,Ellis:1990nz,Okada:1990vk}. However, to use this expression beyond 1-loop order, we have to know the values of the SM~couplings $y_t$ and $g_3$ on the right-hand side at $m_S$. (We continue to ignore electroweak radiative effects such as the running of $g$ and $g^\prime$ in the tree-level term, but these corrections are straightforward to include). Experiment provides the SM~couplings most naturally at lower scales. In a numerical study it is a simple matter to run $y_t$ and $g_3$ to $m_S$, because they do not depend on the ({\it a priori} unknown) weak-scale coupling $\lambda$ at 1-loop order, and in any case this coupling is small. In studies that solve the RGEs analytically in perturbation theory, two different conventions are used for the treatment of the running couplings. On one hand, we can simply solve their RGEs perturbatively to some fixed order, e.g.
\begin{align}
\label{eq:pertyt}
y_t(m_S)=y_t(m_t)+\kappa\left(\beta_{y_t}^{(1)}L+\frac{1}{2}\beta_{y_t}^{(1,1)}L^2\right)+\dots\;,\nonumber\\
g_3(m_S)=g_3(m_t)+\kappa\left(\beta_{g_3}^{(1)}L+\frac{1}{2}\beta_{g_3}^{(1,1)}L^2\right)+\dots\;,
\end{align}
where the parameters on the right-hand side are evaluated at $m_t$, and then insert the analytical expression into Eq.~(\ref{eq:simplelambdamt}). The result should be similar to the fixed-order leading-log computation in the full MSSM~with RG scale $Q=m_S$. On the other hand, this formula for $m_h$ is not of fixed order in the SM~couplings evaluated at $m_t$. If we truncate the formula at a fixed order in the $m_t$-scale couplings, the result should be similar to the fixed-order leading-log computation in the full MSSM~with RG scale $Q=m_t$. For this reason, these two approaches to obtaining analytic formulae for $m_h$ from EFT/RG methods are referred to by the corresponding RG scales of the fixed-order computation in the full theory.

\begin{figure}[t]
\centering
\vspace*{1cm}
\includegraphics[width=0.48\textwidth]{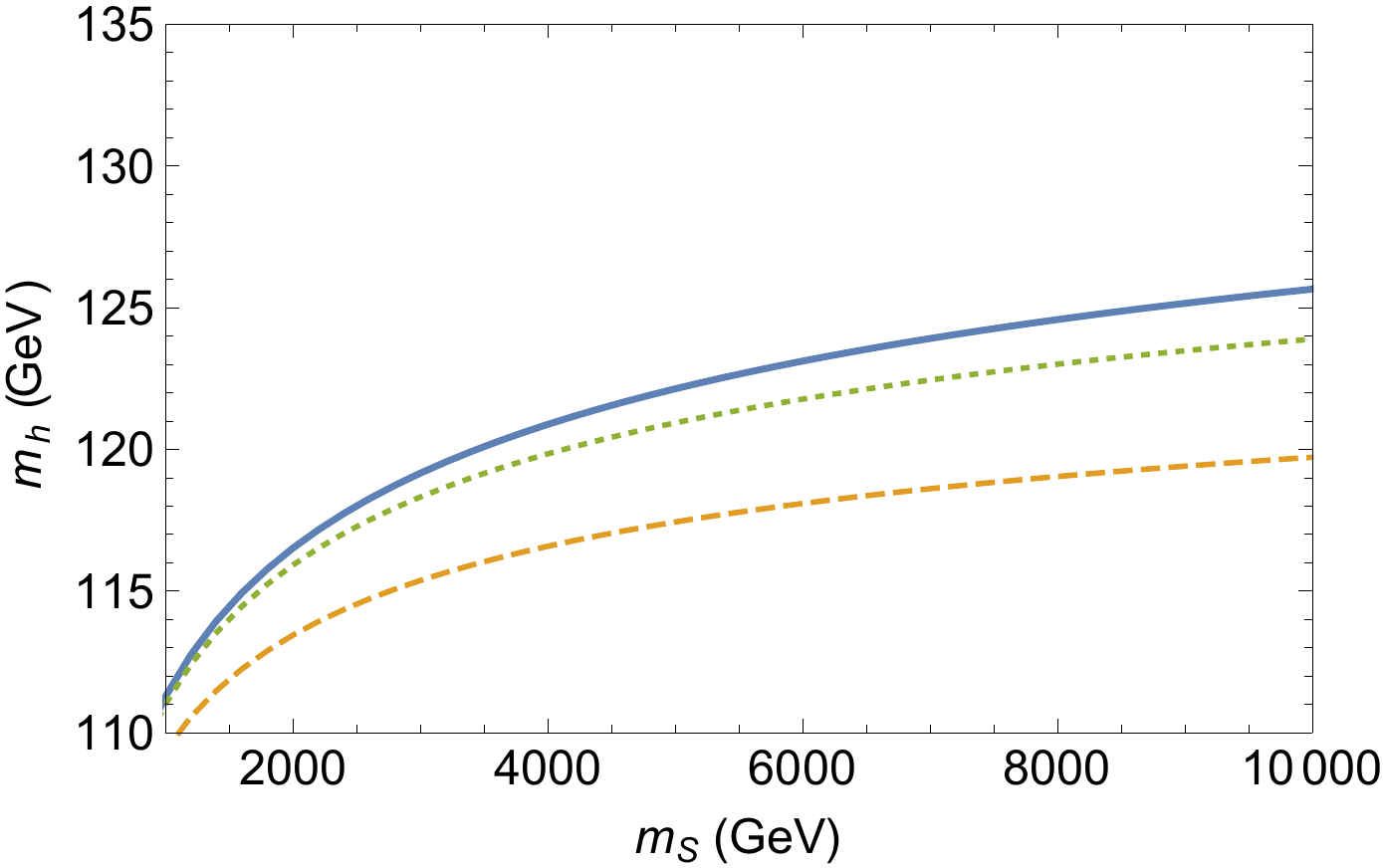} 
\includegraphics[width=0.48\textwidth]{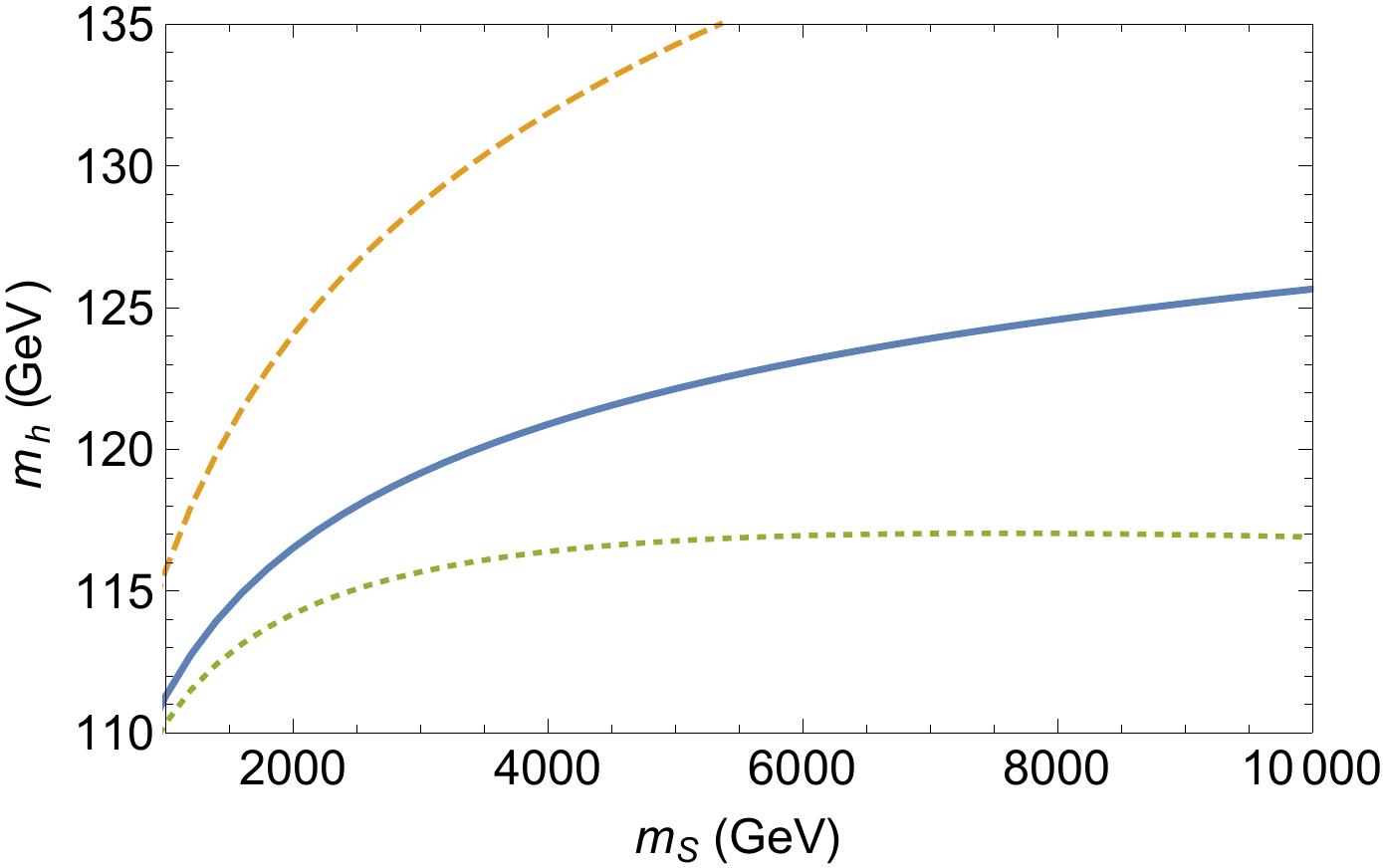} 
\caption{Simple estimates for the Higgs mass $m_h$ as a function of SUSY~scale $m_S$ for $\tb=50$. No threshold corrections are included. Solid curves resum leading logarithmic corrections by numerical solution of the RGEs. In the left panel, the dashed (dotted) curve corresponds to a 1-loop (2-loop) truncation in terms of couplings evaluated at renormalization scale $Q=m_S$. In the right panel, the dashed (dotted) curve corresponds to a 1-loop (2-loop) truncation in terms of couplings evaluated at renormalization scale $Q=m_t$.}
\label{fig:simplemh}
\end{figure}

\quad We can compare the results at different orders in perturbation theory with the more precise ``resummed" result obtained from the numerical integration of the RGEs. In Fig.~\ref{fig:simplemh} we plot the fixed-order and resummed values for $m_h$ as a function of $m_S$ at large $\tb$, where the matching condition for the quartic coupling becomes $\lambda(m_S)=\frac{1}{4}(g^2+g^{\prime 2})$. The two plots show that there can be a significant difference between low-order fixed-order results and the resummed result when $m_S$ becomes large, $\gtrsim 1$ TeV.  The plots also show that for large $m_S$ the fixed-order results through 2-loop order depend strongly on the choice of renormalization scale as described above, and $Q=m_S$ gives the more accurate result. At higher order the curves in both cases must converge to the unique resummed curve. 

\quad Our results so far have captured the most important radiative corrections to the Higgs mass in the MSSM, but they have not been terribly precise. We are missing 1-loop terms in the RGEs (for example, proportional to weak gauge couplings), 1-loop threshold corrections at the weak scale (which determine, for example, the precise relation between the top Yukawa and the top pole mass, and the Higgs quartic and the Higgs pole mass), 1-loop threshold corrections from the decoupling of heavy sparticles at the soft scale $m_S$ (which correct the relations between couplings in the full MSSM~and in the effective theory, and also account for multiple thresholds in the case that the heavy fields are not all exactly degenerate), higher loop corrections of all types (2-loop beta functions, 2-loop threshold corrections, ...), renormalization scheme conversion factors, and higher dimension terms in the Higgs potential generated when the MSSM~is integrated out. It is not conceptually difficult to include most of these corrections, and we review the most important of them in the next few subsections.

\subsection{Threshold and Subleading Logarithmic Corrections}
\begin{figure}[t]
\centering
\includegraphics[width=0.48\textwidth,trim={4cm 17cm 4cm 2cm},clip]{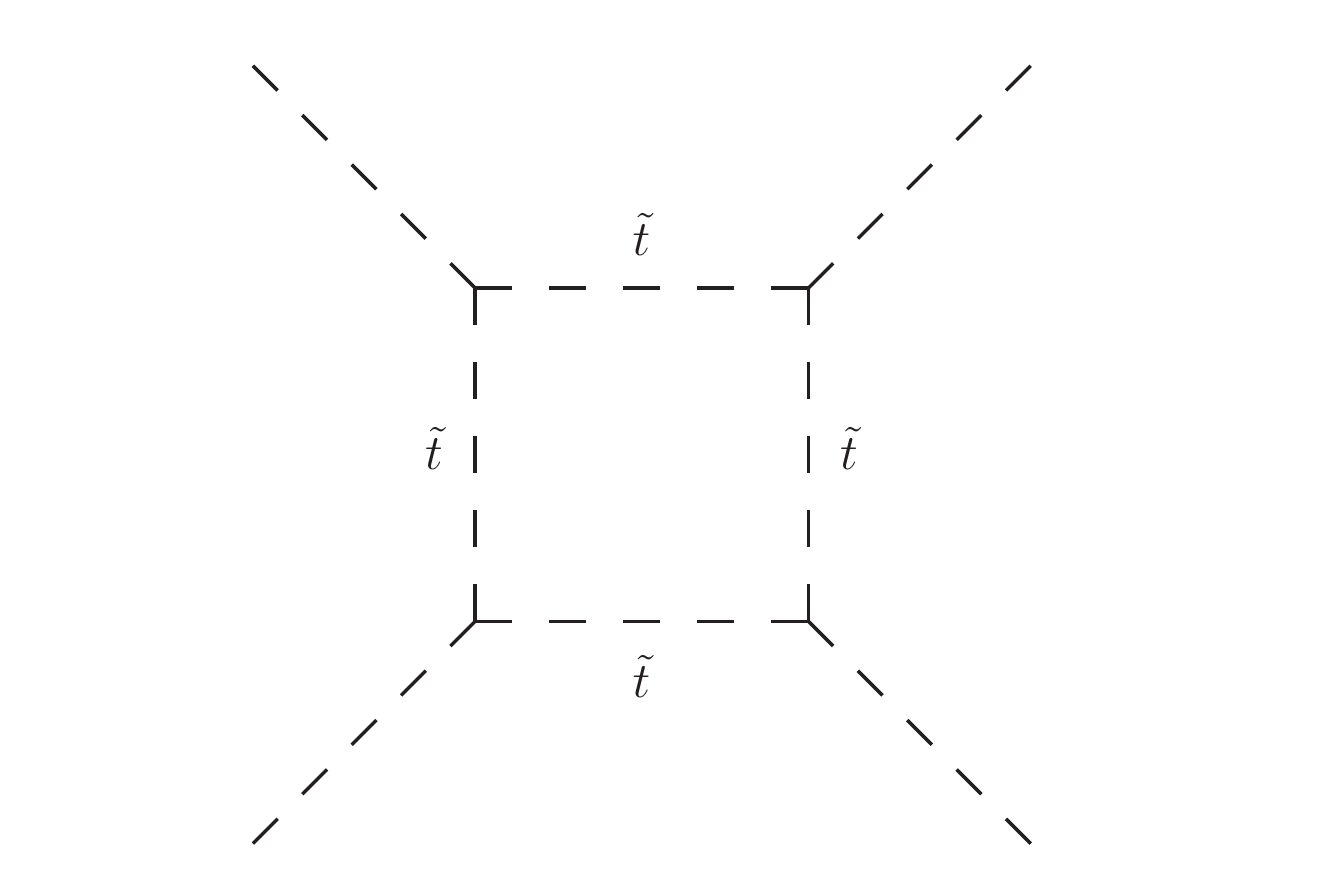} 
\includegraphics[width=0.48\textwidth,trim={4cm 17cm 4cm 2cm},clip]{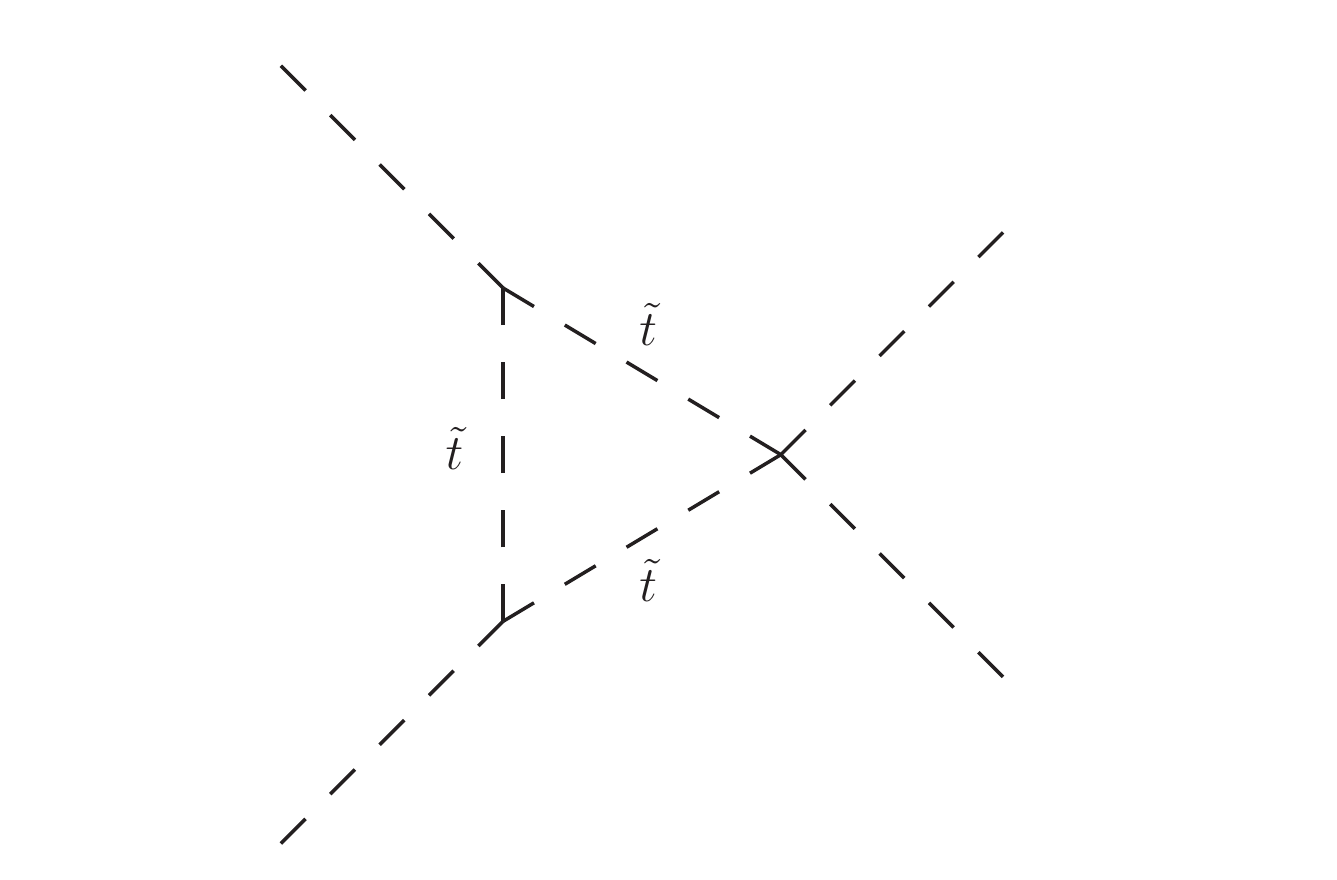} 
\caption{Example 1-loop finite threshold corrections to the Higgs quartic at $m_S$.}
\label{fig:thresholddiag}
\end{figure}

\quad We have already seen how the SM~Higgs quartic coupling at the scale $m_S$ is determined at leading order by the MSSM~electroweak gauge couplings at $m_S$ via tree-level matching. The matching procedure is subject to higher-order corrections from loops involving heavy fields. Through 2-loop order, the leading threshold corrections to $\lambda$ can be organized as:
\begin{align}
\label{eq:lambdaMSSM}
\lambda (M_S) = \lambda^{\text{tree}} & + \D_{\text{th}}^{(\text{sc})} \l + \D_{\text{th}}^{(H,\tilde{N})} \l + \D^{(\a_t)}_{\text{th}} \l + \D^{(\a_b)}_{\text{th}} \l + \D^{(\a_\tau)}_{\text{th}} \l \nl
& + \D^{(\a_s \a_t)}_{\text{th}} \l  + \D^{(\a_t^2)}_{\text{th}} \l . 
\end{align}
In the first line of Eq.~(\ref{eq:lambdaMSSM}), the corrections to the tree level result come from 1-loop renormalization scheme dependence, loops of heavy electroweak fields like the other Higgs bosons and the neutralinos, and loops of heavy stop, sbottom, and stau scalars, respectively. The second line contains 2-loop corrections arising from heavy stops and gluinos as well as scheme conversion effects in the 1-loop corrections.

\begin{figure}[t]
\centering
\includegraphics[width=0.48\textwidth]{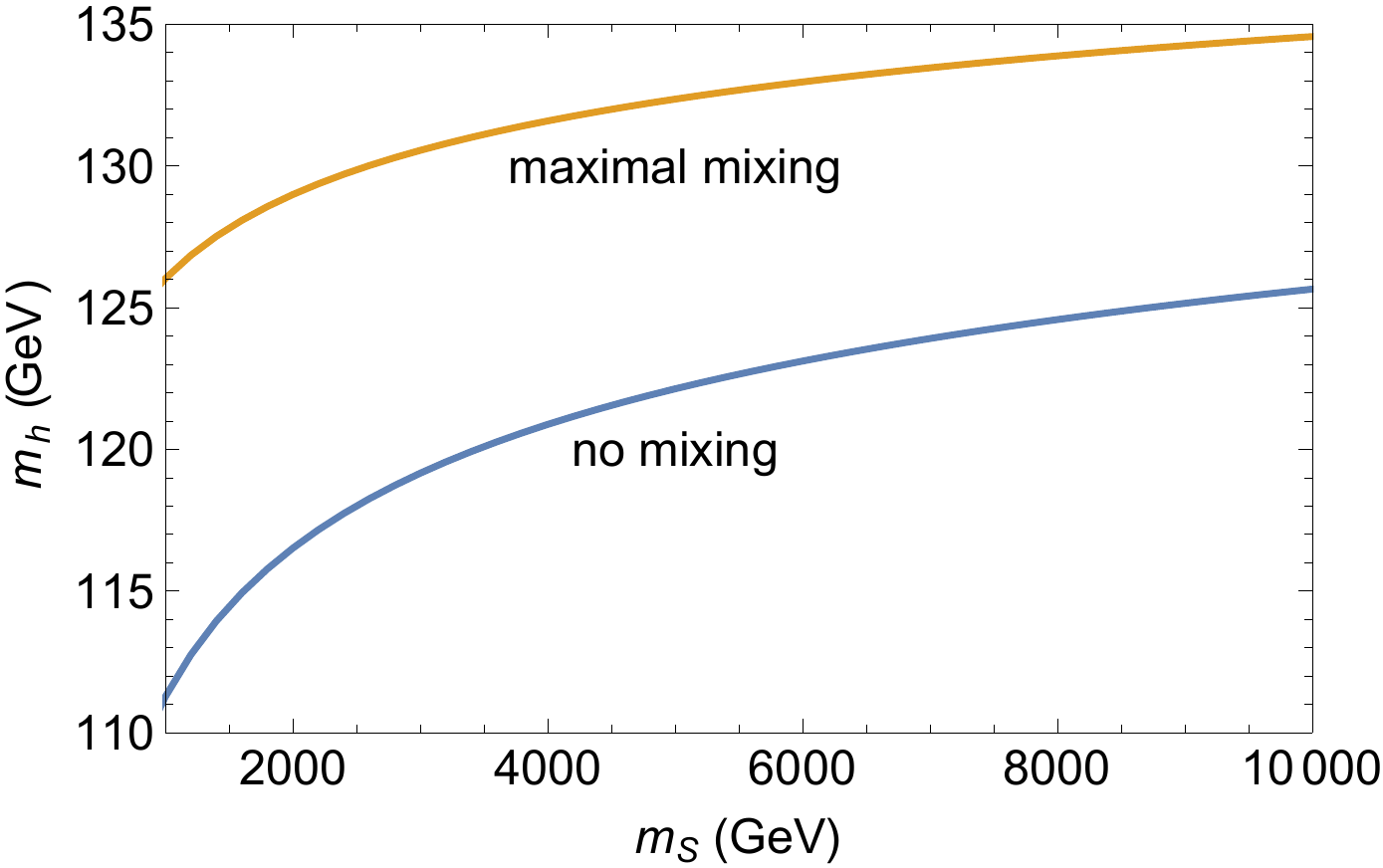} 
\includegraphics[width=0.48\textwidth]{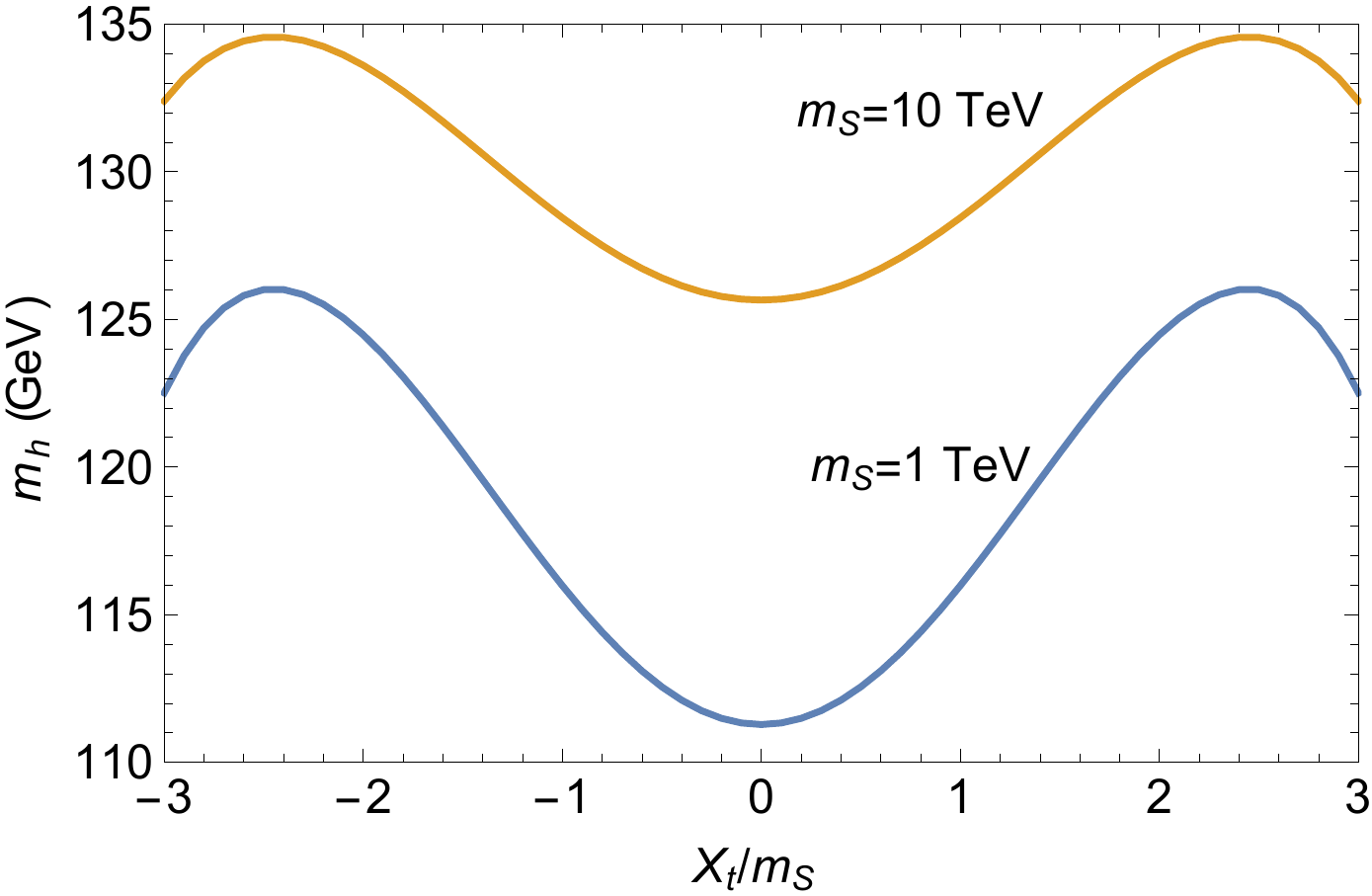} 
\caption{Left: the effect of the 1-loop threshold correction~(\ref{eq:Xtcorr}) on the Higgs mass, computed as in Fig.~\ref{fig:simplemh} with leading-log resummation and $\tb=50$. The bottom curve corresponds to the solid curves in Fig.~\ref{fig:simplemh}, and the top curve corresponds to maximal effect from the correction. Right: the behavior of $m_h$ for fixed $m_S$ and varying stop mixing parameter.}
\label{fig:mixing}
\end{figure}

\quad Of these corrections, $\D^{(\a_t)}_{\text{th}} \l $ has the largest potential to impact $m_h$, and is generated by the diagrams of Fig.~\ref{fig:thresholddiag}~\cite{Carena:1995bx}. These terms are controlled by the stop-stop-Higgs trilinear couplings $X_t=A_t-\mu/\tb$:
\begin{align}
\label{eq:Xtcorr}
\lambda(m_S)=\lambda_{tree}(m_S)+ 6\kappa h_t^4\sinb^4 {\hat X}_t^2(1-{\hat X}_t^2/12)\;,
\end{align}
where $\hat{X}_t=X_t/m_S$, and we have included only the leading term proportional to $h_t^4$. This ``stop mixing" correction is largest at the ``maximal mixing" point, where $\hat{X}_t=\sqrt{6}$. When $m_S$ is of order 1 TeV, $m_h$ can increase by more than 10 GeV when $\hat{X}_t$ is taken from 0 to maximal mixing. This large threshold effect is necessary to achieve $m_h=125$ GeV in the MSSM with TeV-scale superpartners. As $m_S$ becomes larger, the threshold correction's relative impact on $m_h$ decreases, as does its importance since the large logarithms become sufficient to reach 125 GeV. These behaviors are illustrated in Fig.~\ref{fig:mixing}.

\quad Although the exact prediction for the Higgs mass is independent of the renormalization scheme, different schemes organize the perturbative series in different ways. The most common mass-independent scheme used for SM calculations is $\MSbar$ (dimensional regularization with modified minimal subtraction). However, because this scheme breaks supersymmetry, it is customary to use the SUSY-preserving $\DRbar$ scheme (dimensional reduction with modified minimal subtraction) for calculations in the MSSM. If $\MSbar$ parameters are used in the calculation of the Higgs mass, the supersymmetric relation between the MSSM Higgs quartic coupling and the gauge couplings appearing in Eq.~(\ref{eq:treematch}) is modified at 1-loop order~\cite{Haber:1999bp,Giudice:2011cg}:
\begin{align}
\D_{\text{th}}^{(\text{sc})} \l=-\frac{g^4}{64\pi^2 m_W^4}\left(m_Z^4+(2-\frac{2}{3}c^2_{2\beta})m_W^4\right)\;.
\end{align}
Similarly, scheme dependence in the parameters appearing in the 1-loop corrections appears as differences in the 2-loop expressions. For example, for conversions relating on-shell results to mass-independent schemes, see~\cite{Carena:2000dp,Vega:2015fna}. 

\quad Beyond the leading 1-loop threshold corrections in Eq.~(\ref{eq:Xtcorr}), 2-loop threshold corrections are important to obtain an accurate prediction for $m_h$. Expressions for the leading 2-loop corrections to $\lambda$ controlled by the strong gauge coupling and the top Yukawa in the $\DRbar$ scheme may be obtained from the effective potential calculation of~\cite{Espinosa:1999zm,Espinosa:2000df}. If the threshold corrections are expressed in terms of SM $\MSbar$ couplings, as in~\cite{Bagnaschi:2014rsa}, there are further 2-loop corrections to $\lambda$ induced by matching the SM couplings onto the MSSM couplings appearing in 1-loop corrections to $\lambda$. For example, Eq.~(\ref{eq:Xtcorr}) contains the MSSM top Yukawa $h_t$ because the diagrams of Fig.~\ref{fig:thresholddiag} are computed in the full theory. We can obtain the SM top Yukawa at $m_S$ using the RGEs or their perturbative solution~(\ref{eq:pertyt}). The tree-level relation~(\ref{eq:treematch}) relating $y_t$ to $h_t$ (as well as similar relations between $y_{b,\tau}$ and $h_{b,\tau}$) is modified at 1-loop order by squark, gluino, and higgsino loops~\cite{Guasch:2001wv,Carena:2000yi}, resulting in a 2-loop correction to $\lambda$ when expressed in terms of SM couplings. See~\cite{Bagnaschi:2014rsa} for a complete and recent analysis, and~\cite{Kunz:2014gya} for a recent calculation of 2-loop SUSY threshold corrections to the running top Yukawa at $m_S$.

\quad There is another important set of conceptually similar corrections that are implicit in all of our expressions so far. These are corrections to the formulas that determine the SM running couplings from physical observables such as the top quark pole mass. NNLO values for $y_t(m_t)$ and the gauge couplings $g_{1,2,3}(m_t)$ were computed in~\cite{Buttazzo:2013uya} in the $\MSbar$ scheme, and the impact on the MSSM Higgs mass compared with NLO parameters is non-negligible. For example, the 2-loop correction to $y_t$ lowers it by about a percent, which translates into an $\mathcal{O}$(GeV) decrease in $m_h$ for $m_h\sim 125$ GeV. In the other direction, 2-loop corrections can be incorporated that relate the quartic coupling $\lambda(m_t)$ to the Higgs boson pole mass and the mass term in the Higgs potential~\cite{Bezrukov:2012sa,Degrassi:2012ry,Buttazzo:2013uya}.

\quad Subleading logarithmic corrections can also be resummed in an EFT calculation. Perhaps the simplest are the electroweak gauge coupling, bottom Yukawa, and tau Yukawa contributions to the 1-loop beta functions for $\lambda$ and $y_t$. The higher-loop SM beta functions, now known at 2- and 3-loop order~\cite{Arason:1991ic,Luo:2002ey,Buttazzo:2013uya}, can also be implemented.

\quad By concentrating on the matching of renormalizable couplings, we miss contributions to IR physics from higher-dimension operators also generated at $m_S$. One example is the dimension-6 term in the potential, $(H^\dagger H)^3/m_S^2$, generated at $\mathcal{O}(y_t^6)$ by 1 loop of stop squarks. The contribution of higher dimension operators to $m_h^2$ is of order $v^2(v/m_S)^2$ and smaller, and the $(v/m_S)^2$ suppression renders them negligible in the heavy SUSY limit. In contrast, for low $m_S$ these corrections are less suppressed and might be more significant. It is of interest to know the theoretical uncertainty in $m_h$ from the omission of such terms, in particular to inform a choice of whether to use a diagrammatic or EFT calculation (with truncation at dimension-4 operators) for intermediate scales of order $m_S\sim$ few TeV. A simple estimate of the EFT uncertainty from omitting higher dimension operators was performed in Ref.~\cite{Vega:2015fna} by taking the sum of the single-particle corrections to $\Delta\lambda$ and multiplying by $(v/m_S)^2$, and it was found that the error from this source is below a half GeV for $m_S> 1$ TeV. The robustness of this estimate has been questioned~\cite{sventalk}. However, it must be emphasized that in any case the higher dimension operators do not represent an irreducible source of uncertainty: the EFT calculation can be extended to include them in a conceptually straightforward way. Indeed, the derivative-free higher dimension operators were already included at one loop in the calculation of~\cite{Carena:1995wu}, to all orders in $H^\dagger H/m_S^2$, by comparing the 1-loop top/stop correction to $m_h$ obtained from the effective quartic coupling to the correction obtained from the full Coleman-Weinberg effective potential. Including this class of operators, the shift in $m_h$ is typically quite small, less than a few hundred MeV in magnitude for 1 TeV stops and mixing parameter ranging from zero to maximal.

\subsection{State of the Art}

\begin{figure}[t]
\centering
\includegraphics[trim = 0mm -10mm 0mm 0mm, width=0.83\textwidth]{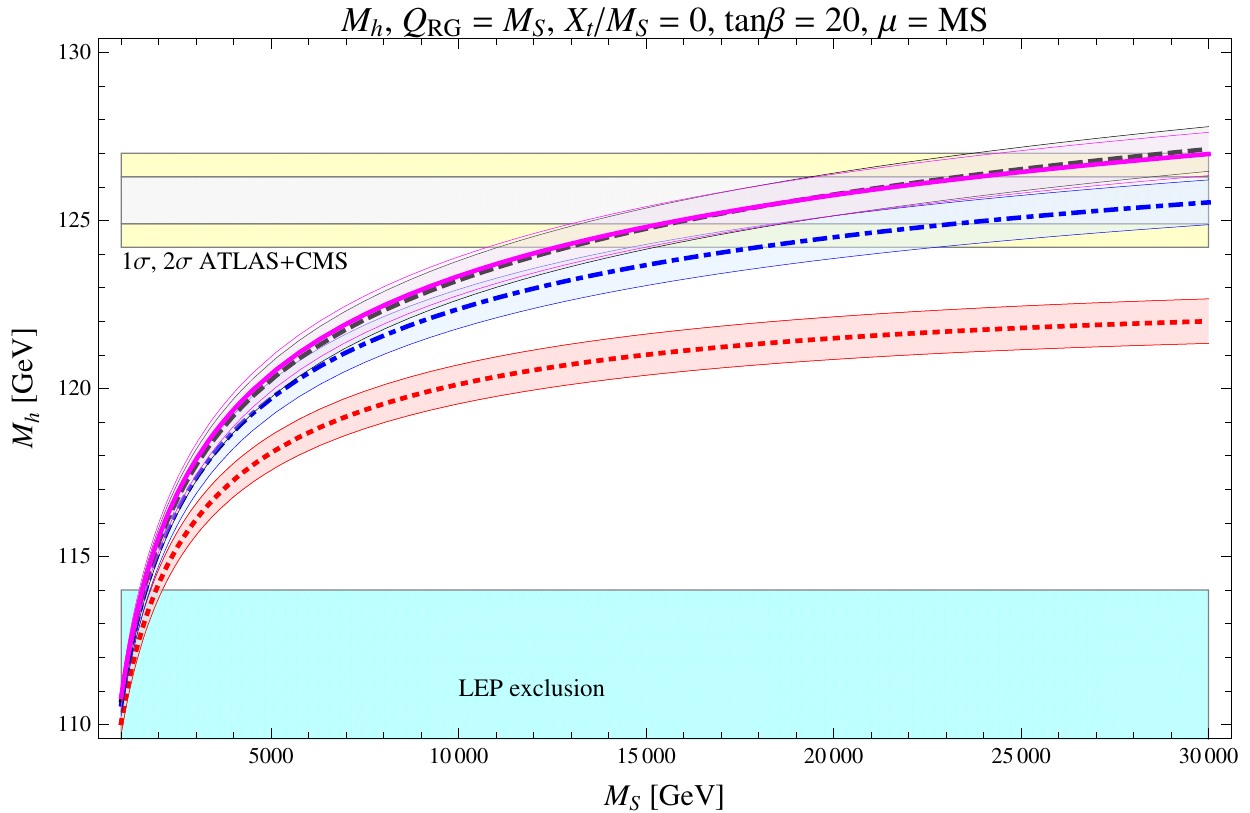} 
\caption{An EFT calculation of $m_h$ in Heavy SUSY showing $m_h$ vs $m_S$ for $\XtMS = 0, \tb = 20$. The magenta (solid) curve is the fully resummed calculation, and the black (dotted), blue (dot-dashed), and red (dotted) curves are the four-, three-, and two-loop fixed-order results obtained by perturbatively solving the RGEs with 3-loop beta functions and $Q=m_S$. Uncertainty bands reflect the variation of $M^{\rm pole}_t$ by $0.7$ GeV. Taken from~\cite{Draper:2013oza}.}
\label{fig:draperetal1}
\end{figure}

\begin{figure}[t]
\centering
\includegraphics[width=0.64\textwidth]{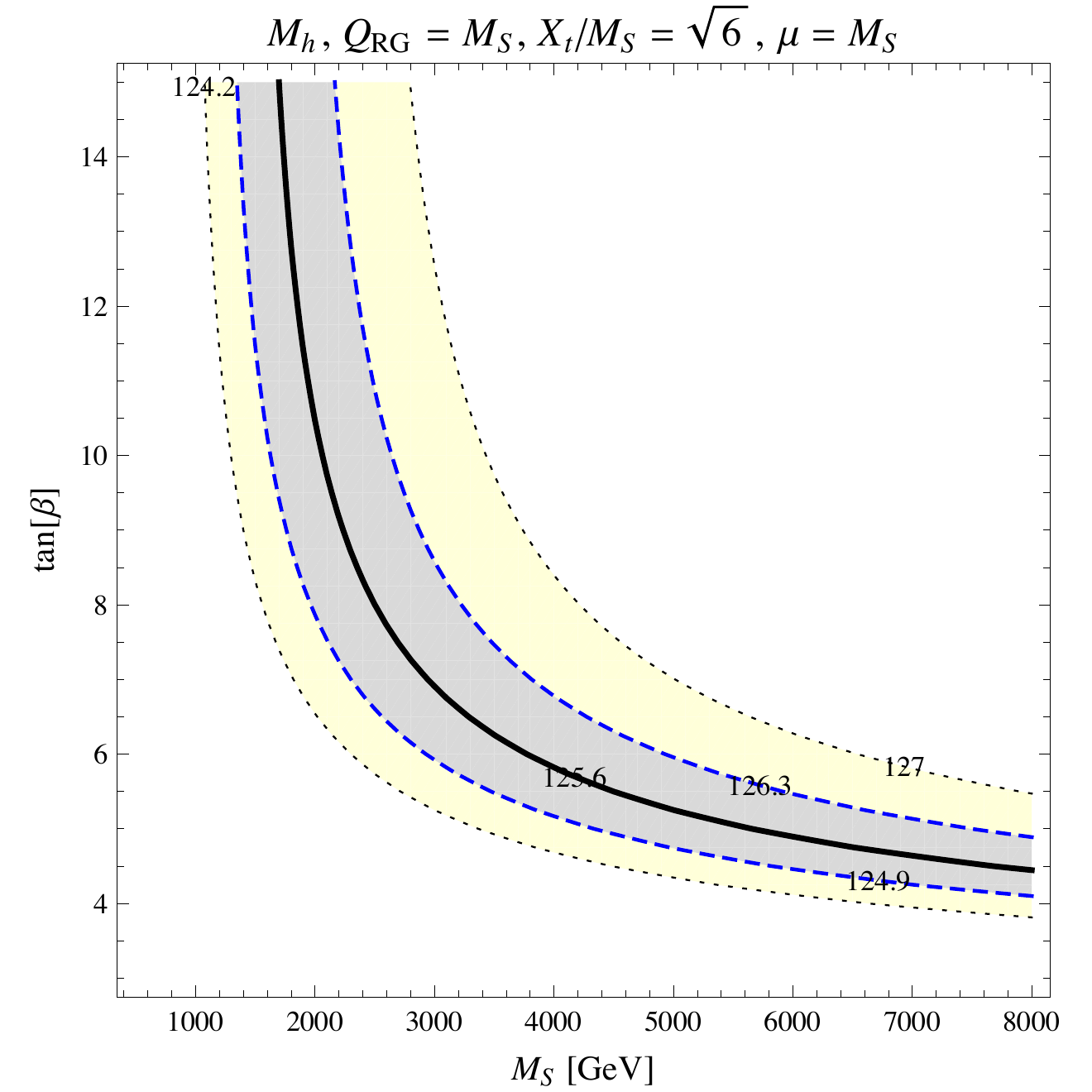} 
\caption{An EFT calculation of $m_h$ in Heavy SUSY showing values of $\tb$ and $m_S$ consistent with $m_h\approx 125$ GeV for $\XtMS = \sqrt6$. Taken from~\cite{Draper:2013oza}.}
\label{fig:draperetal2}
\end{figure}

The most accurate EFT analyses of $m_h$ in the MSSM at present were performed in~\cite{Draper:2013oza,Bagnaschi:2014rsa,Vega:2015fna}. In addition,~\cite{Hahn:2013ria} performed a ``hybrid" calculation, resumming large leading and next-to-leading log terms proportional to the top Yukawa and strong gauge coupling, and including other corrections with a fixed-order Feynman diagrammatic computation. A goal of the hybrid approach is a precise calculation of the Higgs mass
over an intermediate range of scales that may not be efficiently covered by fixed-order or EFT alone.

Despite small differences in the included threshold corrections,~\cite{Draper:2013oza},~\cite{Bagnaschi:2014rsa}, and~\cite{Vega:2015fna} are in close agreement within remaining theoretical uncertainties, estimated in~\cite{Bagnaschi:2014rsa} to be of order 1 GeV at $m_S$ of order 10 TeV. The hybrid calculation of~\cite{Hahn:2013ria} is also in reasonably good agreement at these large scales, with minor unresolved differences suggested to be due to the precision of the top Yukawa calculation and electroweak contributions to $\beta_{y_t}$~\cite{Draper:2013oza,Bagnaschi:2014rsa,pietrotalk,Vega:2015fna}. 

In Figs.~\ref{fig:draperetal1} and~\ref{fig:draperetal2} we reproduce the Heavy SUSY~results of~\cite{Draper:2013oza} for $\XtMS=0$ and $\XtMS=\sqrt6$. In the former case, it is found that stop masses above 10 TeV are required to lift $m_h$ to the experimentally allowed range. In the latter case, a lower SUSY~scale is allowed, of order 2 TeV in the large $\tb$ limit. Compatible results were obtained in~\cite{Bagnaschi:2014rsa,Vega:2015fna}, with~\cite{Bagnaschi:2014rsa} finding agreement to less than half a GeV in a typical parameter point with small stop mixing. 

Few-GeV discrepancies between the hybrid and full EFT approaches in the low-mass range are not yet understood and are under active study. One possibility again relates to the treatment of the top Yukawa extraction~\cite{Draper:2013oza,pietrotalk,Vega:2015fna}, although this has been challenged in~\cite{sventalk}, and it was also suggested in~\cite{sventalk} that the EFT theoretical uncertainties may have been underestimated at low $m_S$.

Analytic fixed-order formulae for $m_h$ from perturbative solutions to the SM RGEs were obtained through 3-loop NLL order in~\cite{Martin:2007pg} and 4-loop NNLL order in~\cite{Draper:2013oza}. The latter was found to provide a good estimate for the complete numerical solution for stop masses up to a few tens of TeV.\footnote{An accidental partial cancellation appears between fixed-order terms of order $\alpha_s^2\alpha_t$ and $\alpha_s \alpha_t^2$ obtained in the EFT approach, first noted in~\cite{Martin:2007pg}. At present, in the Feynman diagrammatic calculation, only terms corresponding to order $\alpha_s^2\alpha_t$ are known~\cite{Harlander:2008ju,Kant:2010tf}; the EFT result may motivate the full diagrammatic calculation of the $\alpha_s  \alpha_t^2$ terms.}

\subsection{EFT Beyond Heavy SUSY}

Single-scale decoupling of the MSSM degrees of freedom may be a poor approximation if some MSSM fields have masses of order the weak scale or if there is more than one large hierarchy in the spectrum. For example, flavor constraints and model-building considerations motivate the possibility that the gauginos are much lighter than the sfermions~\cite{Randall:1998uk,Giudice:1998xp,Wells:2004di,ArkaniHamed:2004fb}. Another possibility with interesting phenomenology is $m_A\sim m_h\ll m_{S}$, in which case the appropriate effective theory below $m_S$ is a 2HDM.

For fixed values of the scalar masses, the Higgs mass in split scenarios can be enhanced over the corresponding mass in heavy SUSY if $\mu,M_{1,2}\ll m_S$~\cite{Arvanitaki:2004eu,Arvanitaki:2012ps,ArkaniHamed:2012gw}. The leading effect is a 1-loop box diagram of higgsinos and winos/binos that contributes a new term to the running of the Higgs quartic below $m_S$,
\begin{align}
\beta_{\lambda,split}^{(1)}=\beta_{\lambda,SM}^{(1)}+\left[6\lambda(g^2+\frac{1}{3}g^{\prime 2})-(g^2+g^{\prime 2})^2-4g^4(1-2s_\beta^2 c_\beta^2)\right]\;.
\label{eq:splitbeta}
\end{align}
The second terms in Eq.~(\ref{eq:splitbeta}) controlled by the electroweak gauge couplings decouple below the scale ${\rm max}(\mu,M_{i})$, but even a 1-loop splitting between scalars and electroweakinos is sufficient to raise the Higgs mass by several GeV.
The most recent high-precision study of the Higgs mass in the presence of split SUSY spectra was performed in~\cite{Vega:2015fna}. It was shown in~\cite{Vega:2015fna} that a full EFT treatment, running with two sets of beta functions above and below the electroweakino mass scale, is in fact very well approximated by the heavy SUSY EFT calculation (with single scale decoupling) supplemented by a one-loop fixed-order correction (essentially including the new terms in Eq.~(\ref{eq:splitbeta}) via the 1-loop term in Eq.~(\ref{eq:lambdall})).

The full EFT treatment of a 2HDM matched onto the MSSM is more involved. The first comprehensive 1-loop study was performed in~\cite{Haber:1993an} and a more recent analysis was performed in~\cite{Cheung:2014hya}. A high-precision study was recently presented in~\cite{Lee:2015uza}. In particular, lower bounds are found on combinations $m_A$ and $\tan\beta$, corresponding to parameter points at which the suppression of $m_h$ by mixing effects is so large that radiative corrections to $m_h$ from stops and electroweakinos cannot accommodate $m_h=125$ GeV.  We refer the reader to~\cite{Lee:2015uza} for further details.

\section{Public Computer Programs}
\label{sec:surv-publ-comp}
\begin{table}[t]
\label{tbl:codes}
\begin{tabular}{|l|c|c|c|c|c|c|c|}
\hline
                     &  fixed order                                  &                 RL                                     &  scheme                 &        CPV?         &        RPV?      &   BMSSM?  \\
\hline
SUSPECT~\cite{Djouadi:2002ze}    &         2L                                         &               ---                                        & $\DRbar$              &           ---            &          ---          &        (gen.)         \\ 
SPheno~\cite{Porod:2003um,Porod:2011nf}        &         2L                                        &               ---                                        & $\DRbar$               &  \checkmark     &  \checkmark     &       gen.       \\
SoftSUSY~\cite{Allanach:2001kg,Allanach:2013kza,Allanach:2014nba}     &         2L                                        &               ---                                        & $\DRbar$               &           ---            &  \checkmark   &    NMSSM   \\ 
CPsuperH~\cite{Lee:2003nta,Lee:2007gn,Lee:2012wa}    &         2L                                        &               ---                                       & $\MSbar$                 &   \checkmark     &         ---           &        ---         \\
FeynHiggs~\cite{Heinemeyer:1998yj,Heinemeyer:1998np,Degrassi:2002fi,Frank:2006yh,Hahn:2013ria}    &         2L                                        &              2L                                       &   mixed                   &  \checkmark      &         ---          &    (NMSSM) \\
SUSYHD~\cite{Vega:2015fna}      &        ---                                         &               3L                                       & $\DRbar$                & \checkmark       &          ---          &         ---         \\
H3m~\cite{Kant:2010tf}             &         3L                                        &                ---                                        & $\DRbar$ /mixed   & ---                      & ---                   &         ---          \\
NMSSMTools~\cite{Ellwanger:2004xm,Ellwanger:2005dv} &        2L                                       &                 ---                                        &$\DRbar$               & (\checkmark)      &  ---                   &    NMSSM    \\
NMSSMCalc~\cite{Baglio:2013iia}     &     2L                                       &                 ---                                        &$\DRbar$/mixed     & \checkmark       &         ---            &      NMSSM   \\
FlexibleSUSY~\cite{Athron:2014yba} &      2L                                        &                 ---                                        &$\DRbar$               & ---                      &       ---              &      gen.        \\
\hline
\end{tabular}
\caption{\label{tab:codesummary} Summary of public codes. The column ``fixed order" denotes the loop order to which the Higgs spectrum is calculated. In this column ``2L" typically indicates that the known two-loop corrections have been incorporated, while the code H3m includes also the three-loop corrections at order $\alpha_s^2\alpha_t$. CPsuperH is fixed
  order in that sense that though the effective 
  field theory approach has been applied, the resummation of the leading and
  next-to leading logarithms is
  truncated at 2-loop order. The column ``RL" indicates the loop order at which large logarithmic corrections are resummed; FeynHiggs uses two-loop beta functions for $y_t$ and $g_3$, while SUSYHD uses three-loop beta functions for all SM couplings. The columns ``CPV" and
``RPV" indicate whether the codes permit  CP-violating and R-parity violating parameters, and the column ``BMSSM" indicates whether the codes can accommodate models beyond the MSSM.
NMSSMCalc and NMSSMTools are NMSSM-specific codes, while Spheno and
FlexibleSUSY offer the possibility to generate a particle mass spectrum for
more general models linking the computer code Sarah, which is indicated by
``gen.''. Parenthesis indicate features that are in progress but not yet publicly available.}
\end{table}

\quad A number of groups have created publicly-available codes for precision computations in supersymmetric models. Together, the codes calculate a broad range of phenomenological properties, including spectra, branching ratios, collider cross sections, low-energy observables, renormalization group behavior, and cosmological predictions. Originally, most codes only performed calculations in simple limits of the MSSM. Subsequently the codes underwent sophisticated development to operate on the full MSSM including flavor, $CP$, and $R$-parity violating effects, and more recently they have evolved to calculate in a wide variety of beyond-the-MSSM (BMSSM) models. Here we briefly review the most important features of the Higgs mass calculations implemented by the various codes in the context of the MSSM. A summary is given in Table~\ref{tbl:codes}.

\quad Most public codes, with two recent exceptions discussed below, perform what are in essence fixed-order computations of $m_h$ in the full MSSM. The most sophisticated fixed-order codes, CPsuperH~\cite{Lee:2003nta,Lee:2007gn,Lee:2012wa}, SPheno~\cite{Porod:2003um,Porod:2011nf}, SuSpect~\cite{Djouadi:2002ze}, SoftSUSY~\cite{Allanach:2001kg,Allanach:2013kza,Allanach:2014nba}, FeynHiggs~\cite{Heinemeyer:1998yj,Heinemeyer:1998np,Degrassi:2002fi,Frank:2006yh,Hahn:2013ria}, and H3m~\cite{Kant:2010tf}, calculate full 1-loop radiative corrections to the Higgs spectrum as well as the leading 2-loop corrections controlled by the strong gauge coupling and third-generation Yukawa couplings, and in the case of H3m, 3-loop corrections at order $\alpha_s^2\alpha_t$. These calculations are expected to be highly accurate for superpartners near the 1 TeV scale, where $\log(m_S/m_t)$ is not large enough to require EFT techniques.

\quad The codes differ firstly in their choice of renormalization scheme. In its default configuration, FeynHiggs performs an on-shell calculation. CPsuperH implements an $\MSbar$ calculation, and is most similar to the EFT calculation with 2-loop series expansion around $Q=m_t$~\cite{Carena:2001fw}. SPheno, SuSpect, and SoftSUSY are the most similar to each other, using the $\DRbar$ scheme with user-controlled choice of renormalization scale. All three numerically solve 2-loop RGEs to run couplings and masses to the chosen scale, and SoftSUSY has recently implemented 3-loop RGEs~\cite{Allanach:2014nba}. H3m is also a $\DRbar$ calculation, converting the 2-loop on-shell result from FeynHiggs to $\DRbar$ and adding the $\alpha_s^2\alpha_t$ 3-loop correction.

\quad Among the 2-loop $\DRbar$ codes, small differences arise in the computations of the threshold corrections that relate SM observables to MSSM running couplings. The most numerically significant differences are formally of higher order in the perturbative expansion, and therefore no one prescription is clearly preferred. The spread in the results for $m_h$ can then be interpreted as a measure of the theoretical uncertainty from missing higher-order corrections. Early comparisons between SPheno, SuSpect, and SoftSUSY were performed in~\cite{Allanach:2004rh}. 

\quad At present, FeynHiggs and the code SUSYHD~\cite{Vega:2015fna} are the only public codes that can perform an EFT calculation. The EFT implementation in FeynHiggs uses a hybrid scheme to resum next-to-leading logarithmic terms controlled by the strong gauge couplings and the top Yukawa and add them to the fixed-order subleading terms from a Feynman diagrammatic calculation~\cite{Hahn:2013ria}. SUSYHD is a Mathematica package offering a full EFT calculation (3 loop SM running + 2 loop matching) for both heavy SUSY and split SUSY.

\subsection{Existing Comparisons of Fixed-Order and EFT Codes}

\begin{figure}[t]
\centering
\includegraphics[trim = 10mm 0mm 0mm 0mm, width=0.53\textwidth]{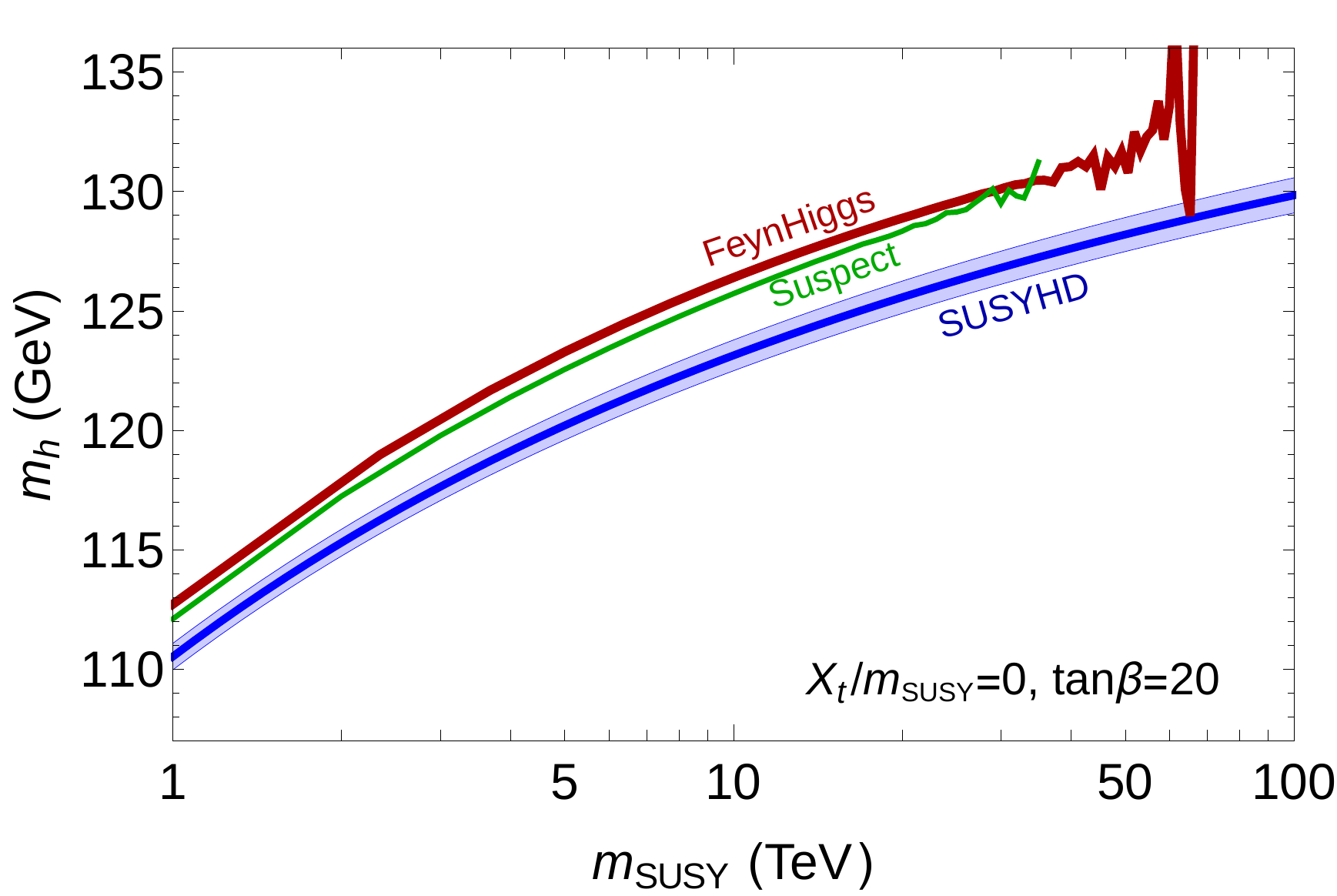} 
\caption{Comparison of the fixed-order code Suspect, the hybrid code FeynHiggs, and the EFT code SUSYHD as a function of the soft SUSY-breaking scale, taken from~\cite{Vega:2015fna}. Above 20 TeV, Suspect encounters instabilities; FeynHiggs is stable up to scales of order 40 TeV; and the fully-resummed calculation in SUSYHD allows calculations to much higher scales. However, the sources of discrepancies in regions where all codes are stable, particularly at scales of order 1 TeV, and the robustness of theoretical uncertainties in the EFT calculation at such scales, is not widely agreed-upon at present~\cite{Bagnaschi:2014rsa,Vega:2015fna,sventalk}.}
\label{fig:vegaetal}
\end{figure}

\quad With both fixed-order and EFT public codes available, it is clearly of interest to know in which regimes of parameters, particularly $m_S$, one type of calculation is favored over the other.  At present, this question has only a qualitative answer without a sharp boundary, and is an active area of study. Here we briefly summarize some of the recent results.

\quad Ref.~\cite{Bagnaschi:2014rsa} compared modern versions of SPheno, SuSpect, SoftSUSY, and FeynHiggs to each other and to an EFT calculation resumming next-to-next-to-leading logarithmic terms. For 1 TeV superpartners, as discussed above, the existing EFT calculations are missing $v/m_S$ terms that are not highly suppressed. Relative to the EFT, the 2-loop fixed order codes primarily miss higher-order logarithmic terms, although the logs are not large in this case, and higher-order terms associated with the extraction of $h_t$ from $m_{t,pole}$. Ref.~\cite{Bagnaschi:2014rsa} found that for large $A_t$ (necessary to achieve $m_h\simeq125$ GeV in the MSSM with TeV-scale SUSY), there is a 2 GeV spread among $\DRbar$ fixed-order codes around 125 GeV, the EFT calculation gives an $m_h$ more than a GeV lower than those codes, and FeynHiggs yields a result several GeV higher. The large spread suggests that higher order corrections will still contribute at least $\pm3$ GeV to some of the calculations for $m_S$ of order 1 TeV. 

\quad For superpartners heavier than 2-3 TeV, the 2- and 3-loop fixed-order computations employed by current codes are expected to lose precision due to the absence of logarithmically-enhanced higher-order corrections. For example, at $m_S=3$ TeV, running FeynHiggs in its fixed order vs hybrid modes shifts $m_h$ by $\sim 3$ GeV (although the magnitude of this shift depends on the renormalization scale chosen in the fixed-order computation, and can be made smaller by choosing renormalization scales of order $m_S$ instead of $m_t$, due to the decrease in $h_t$).  In both the FeynHiggs hybrid mode and in SUSYHD, the result for $m_h$ is more stable for $m_S$ above 10 TeV than in purely fixed-order calculations. This is exhibited in Fig.~\ref{fig:vegaetal}, taken from~\cite{Vega:2015fna}. At large scales the FeynHiggs result remains typically slightly larger than what is found with SUSYHD and other EFT calculations, and in particular is currently outside the large-$m_S$ theoretical uncertainty estimated in~\cite{Bagnaschi:2014rsa}. As mentioned previously, this may be due to the use of the NLO top Yukawa in~\cite{Hahn:2013ria} (consistent with the use of 2-loop beta functions in FeynHiggs) vs. the NNLO+partial N$^3$LO value used in the other calculations.

\subsection{Beyond the MSSM}

\quad Recently, many codes have developed for precision calculations in models beyond the MSSM. NMSSMTools~\cite{Ellwanger:2004xm,Ellwanger:2005dv} computes the Higgs masses and decay rates in the NMSSM including leading two-loop corrections to the spectrum. NMSSMCALC~\cite{Baglio:2013iia} also computes spectra and phenomenology in the NMSSM, and in addition allows explicit CP violation in the parameters.\footnote{NMSSMTools is also being extended to include complex parameters; see~\cite{Domingo:2015qaa}.}
SoftSUSY has likewise been extended to the NMSSM~\cite{Allanach:2013kza} and linked to NMSSMTools~\cite{Ellwanger:2004xm,Ellwanger:2005dv}, and SuSpect has also recently been expanded to permit implementation of BMSSM calculations~\cite{suspect3}. A particularly versatile tool is the ``spectrum generator generator" SARAH~\cite{Staub:2008uz,Staub:2009bi,Staub:2010jh,Goodsell:2014bna,Staub:2013tta}, which can convert Lagrangian input into spectrum-calculating source code for injection into SPheno~\cite{Porod:2003um,Porod:2011nf} and FlexibleSUSY~\cite{Athron:2014yba}. In the Higgs sector, both Spheno and FlexibleSUSY calculate full 1-loop radiative corrections, while Spheno can be extended to include the dominant two-loop corrections in the effective potential approximation~\cite{Goodsell:2014bna}, and FlexibleSUSY can compute leading two-loop corrections in the MSSM and NMSSM.  A recent comparison of NMSSM codes was performed in~\cite{Staub:2015aea}, and differences between codes were traced primarily to the extractions of $\DRbar$ parameters from SM observables and types of 2-loop corrections included in the calculation.

\section{Conclusions}

The mass of the Higgs boson is a sensitive probe of physics beyond the Standard Model. We have reviewed the precision calculation of the lightest Higgs mass in the Minimal Supersymmetric Standard Model, where $m_h$ receives large radiative corrections sensitive to a variety of other masses and couplings. Two methods of calculation stand out, with different strengths and weaknesses: the Feynman-diagrammatic calculation, capturing all radiative corrections order by order in a loop expansion, and the effective field theory calculation, which captures corrections with large logarithms to all orders. We have described in detail the calculation of the simplest leading terms in both cases and explained the sources of higher-order corrections. We have also summarized the state of the art for both calculations and reviewed the public codes available to compute the Higgs mass numerically over the MSSM parameter space.

Of particular interest to the LHC program is the possibility that (some) superpartners lie below $\sim1-2$ TeV. Which experimentally accessible regimes of MSSM parameters, in particular the stop masses and mixings and $\tan\beta$, are compatible with $m_h=125$ GeV? At present, the Feynman-diagrammatic and effective field theory calculations yield different answers to this important question, pointing to theoretical uncertainties that can only be reduced with further calculation of higher-order corrections. Natural next steps for the Feynman-diagrammatic approach include the calculation of additional 3-loop terms at fixed order, as well as the inclusion of 2-loop threshold corrections and 3-loop beta functions in the hybrid calculation. The uncertainties in the EFT approach can be reduced by extending the modern calculations to include the matching of higher dimension operators. We hope that this review will provide a useful entry point to researchers interested in contributing to this timely program.
\label{sec:concl}

\section*{Acknowledgement}
We would like to thank Jonathan Feng for suggesting this review to us. We also thank Philipp Kant for collaboration during early stages of this work, Luminita Mihaila for helpful assistance with H3m and Fig. 3, Gabriel Lee and Carlos Wagner for conversations, and our referee for valuable suggestions. This work is partially supported by the Danish National Research Foundation under grant DNRF:90.

\bibliography{review-mh}
\bibliographystyle{elsarticle-num}
\end{document}